\documentclass[aps,pre,reprint,amssymb,superscriptaddress]{revtex4-1}

\usepackage{bm}
\usepackage{graphicx}
\usepackage{amsmath}
\usepackage{amsthm}
\usepackage{eucal}
\usepackage{amssymb}
\usepackage{mathrsfs}
\usepackage{enumerate}
\usepackage{xcolor}
\usepackage{array}

\newcommand{\comments}[1]{}

\begin{document}

\title{A mapping of the stochastic Lotka-Volterra model to models of population genetics and game theory}

\author{George W.~A.~Constable}
\affiliation{Department of Evolutionary Biology and Environmental Studies,
University of Zurich, 8006 Zurich, Switzerland}
\author{Alan J.~McKane}
\affiliation{Theoretical Physics Division, School of Physics and Astronomy,
The University of Manchester, Manchester M13 9PL, United Kingdom}

\begin{abstract}

The relationship between the $M$-species stochastic Lotka-Volterra competition (SLVC) model and the $M$-allele Moran model of population genetics is explored via timescale separation arguments. When selection for species is weak and the population size is large but finite, precise conditions are determined for the stochastic dynamics of the SLVC model to be mappable to the neutral Moran model, the Moran model with frequency-independent selection and the Moran model with frequency-dependent selection (equivalently a game-theoretic formulation of the Moran model). We demonstrate how these mappings can be used to calculate extinction probabilities and the times until a species' extinction in the SLVC model. 
\end{abstract}
\maketitle

\section{Introduction}
\label{sec:intro}

Perhaps the most important models in ecology, population genetics and game theory are respectively the generalized Lotka-Volterra (LV) model~\cite{roughgarden_1979}, the Moran model~\cite{moran_1957} (and its discrete-generation variant the Wright-Fisher model~\cite{fisher_1930}), and the replicator equations~\cite{hofbauer_1998}. The generalized LV model describes the dynamics of an arbitrary number of species interacting in a pairwise fashion according to an interaction matrix (which can be used to describe competitive, mutualistic and predatory interactions), and is almost always treated deterministically~\cite{pielou_1977}. The Moran model describes the evolution of a population of individuals carrying different alleles in a way that accounts for genetic drift, and is therefore inherently stochastic~\cite{halliburton_2004}. The replicator equations describe the time-evolution of the frequency of players playing a given strategy in a pairwise game, where strategies increase according to the average payoff players receive from that strategy when playing against the population~\cite{hofbauer_1998}. Historically these models were viewed deterministically, however the last decade has seen a surge of interest in incorporating and analyzing stochasticity in these models~\cite{nowak_2004,traulsen_2005,traulsen_2008}, typically through casting this as a birth-death process analogous to the Moran model. 

In this paper, we begin by developing a stochastic analog of the generalized LV model for $M$ species. This stochastic Lotka-Volterra competition (SLVC) model does not have a fixed population size (the number of individuals in the system is free to vary). We then set about determining the conditions under which its stochastic behavior can be seen at long time to be equivalent to the neutral haploid multiallelic Moran model~\cite{kimura_1956}, the haploid multiallelic Moran model with constant selection~\cite{zeng_1989,mallet_2012} (variously termed the Moran model with directional selection~\cite{otto_2008} or frequency-independent selection~\cite{mallet_2012}) and a Moran version of a game-theoretic model of pairwise games with multiple strategies~\cite{antal_2009}. The Moran model in each instance features a population of fixed size $N$ (the number of individuals in the system is fixed). We will show that the SLVC model can be mapped onto each of the above mentioned processes under conditions which we summarize in tabular form. Our analysis relies on timescale separation arguments, and is dependent on the process of population regulation occurring on a much faster timescale than that of the change in population composition. Our results are thus valid when selection between species/alleles/strategies is weak and $N$ is large but finite.

Although the links between the models which we identify have not been discussed previously, other ways in which they are related to each other have been explored, and we will now review these. It is perhaps not surprising that the deterministic versions of these models have been studied far more thoroughly than their stochastic analogs. It is well known that the deterministic LV model in $M$ variables can be mapped to the replicator equations in $M+1$ variables by the inclusion of an additional variable in the LV model that keeps track of population size~\cite{hofbauer_1998}. While this first result by Hofbauer~\cite{hofbauer_1981} allowed for a plethora of results to be obtained, these were all entirely in the deterministic limit (see~\cite{bomze_1995} for a comprehensive review). 

For our purposes, the most relevant work was not until the publication of Ref.~\cite{nowak_2004}, when interest was ignited in incorporating demographic stochasticity into the replicator equations. Here it was first demonstrated that the Moran model could be formulated with reproduction rates that vary with population composition to give analogous dynamics to the replicator equations in the infinite population size and weak selection limit. Since then various choices have been explored for how reproduction rates in the Moran model might be dependent on population composition (e.g. linear~\cite{nowak_2004}, exponential~\cite{traulsen_2008}). These all share the common feature that they become functionally similar in the limit of weak selection and by construction also to the replicator equation in the limit of infinite population size. However, this equivalence between game theoretic formulations of the Moran model and the replicator equations has been shown only to hold in the limit of weak selection~\cite{traulsen_2005}. When selection is strong, alternative update processes, such as an imitation process, map more cleanly to deterministic replicator dynamics~\cite{traulsen_2005}.

More recently, efforts have been made to understand the role of demographic noise in the LV model by developing stochastic LV models~\cite{khasminskii_2001,spagnolo_2003,zelnik_2015}. Quite naturally with this has come a desire to understand how these probabilistic systems might be related to other canonical models. In \cite{noble_hastings_2011} it was shown that a similar frequency-dependent Moran model to that used in the game theory literature shares deterministic fixed points with the LV model. This is perhaps unsurprising, as this formulation of the Moran model by construction maps to the replicator equation in the deterministic limit, which in turn is known to map to the LV model. It is then suggested in \cite{noble_hastings_2011} that this equivalence in the deterministic limit might be used to try to understand the stochastic dynamics of the LV model. However this misses the crucial point that two systems that share the same deterministic dynamics need not have the same stochastic dynamics. This was demonstrated in the present context in Ref.~\cite{parsons_quince_2010}, where it was shown that while increasing the longevity of a type in an SLVC model (while keeping its lifetime fecundity fixed) had no effect on the deterministic dynamics, it could affect the stochastic dynamics, increasing the fixation probability of the type. In that paper the emergent differences between the Moran model with constant selection and a formulation of the SLVC model with homogeneous competition were discussed, primarily for the two-allele case. Finally, the SLVC model in two variables has been mapped to the Moran model in a single variable~\cite{constable_2015}. 

As we have indicated, in this paper we will be using timescale separation arguments in developing a mapping between the SLVC model and three formulations of the Moran model; the neutral model, the model with constant selection and the replicator model. Unlike traditional deterministic approaches that map $M$ LV equations to $M+1$ replicator equations~\cite{hofbauer_1998}, we will map $M$ SLVC equations to $M-1$ stochastic replicator equations of the Moran type by elimination of the fast transient associated with a fast approach of the system to carrying capacity. When dealing with both the Moran models and the SLVC model, we will exclusively look at the limit of weak selection. We will therefore not have to be overly concerned about the breakdown of equivalence of the Moran formulation of the replicator equations and the replicator equations in large selection strength regimes~\cite{traulsen_2005}. In applying our dimensional reduction, we will be careful to deal correctly with the noise terms in the SLVC model. This allows us to determine a full stochastic mapping between the SLVC model and the Moran models, rather than simply inferring the mapping based on a deterministic equivalence~\cite{noble_hastings_2011}. We will not consider noise-induced selection effects of the type identified in \cite{parsons_quince_2010}, as these lie outside the scope of this paper. However unlike in Ref.~\cite{parsons_quince_2010}, where competition rates were taken to be symmetric,we will analyze the effect of varying competition rates. This will allow for the extension of the mapping from the SLVC model to the replicator equations, as well as giving the conditions that the competition matrix must satisfy in order for the Moran model with constant selection to be a valid approximation. Finally, this paper will provide a multivariate extension to the work reported in Ref.~\cite{constable_2015}. Not only does this extend the treatment given there, but the multivariate analysis also provides a deeper insight into the mapping between the models.

\section{Model definitions and the mesoscopic formulation}
\label{sec:model_defn}

We define the models at the very basic level of individuals which are born and die and where changes occur from one type to another due to the process of competition. In these individual-level models or individual-based models, the state of the system at a given time is specified by how many individuals of the different types are present at that time. The models are essentially defined by giving functional forms for the rates at which transitions from one state to another occur. We will now describe these for each of the models in turn, starting with the SLVC model.

\subsection{The SLVC model}
\label{sec:SLVC_model_defn}
As discussed in the Introduction, the system is a population of $n_1$ haploid individuals each of which carries an allele of type $1$, $n_2$ haploid individuals each of which carries an allele of type $2$, ..., $n_M$ haploid individuals each of which carries an allele of type $M$. We denote the state of the system by the vector $\bm{n}=(n_1,\ldots, n_M)$. Individuals of type $\alpha$ reproduce at a rate $b_\alpha$ and die at a rate $d_\alpha$, $\alpha=1,\ldots,M$. The total number of individuals, $\sum^M_{\alpha=1} n_\alpha$ is not fixed, instead it is regulated by the process of competition, which occurs between individuals of type $\alpha$ and $\beta$ at a rate $c_{\alpha \beta}$.

The transition rates from state $\bm{n}$ to a new state $\bm{n}'$ are generalizations of those given for the case of two alleles in Ref.~\cite{constable_2015}:
\begin{eqnarray}\label{trans_rates}
T_{\alpha +}(n_\alpha + 1|n_\alpha) &=& b_\alpha \frac{n_\alpha}{V}, \nonumber \\
T_{\alpha -}(n_\alpha - 1|n_\alpha) &=& d_\alpha\frac{n_\alpha}{V} + \sum^{M}_{\beta = 1} c_{\alpha \beta}\frac{n_\alpha}{V}\frac{n_\beta}{V}, 
\label{trans_rates_SLVC}
\end{eqnarray}
where $\alpha=1,\ldots,M$ and where only the alleles which change in number have been given as arguments of the transition rates (the original state is to the right and the new state to the left). The parameter $V$ is a measure of the size of the system, such as the volume. It will be shortly be used to make a transition to a mesoscopic description, via the diffusion approximation. 

Since the transition rates in Eq.~(\ref{trans_rates_SLVC}) only depend on the current state of the system, the process is Markovian, and can be described by a master equation for the probability, $P_{\bm{n}}(t)$, of finding the system in state $\bm{n}$ at time $t$~\cite{van_Kampen_2007}. It is given by
\begin{eqnarray}
\frac{\mathrm{d}P_{\bm{n}}(t)}{\mathrm{d}t} &=& \sum^{M}_{\alpha=1}
\left[ T_{\alpha +}(\bm{n}|\bm{n}-\bm{\nu}_{\alpha +})
P_{\bm{n}-\bm{\nu}_{\alpha +}}(t) \right. \nonumber \\
&+& \left. T_{\alpha -}(\bm{n}|\bm{n}-\bm{\nu}_{\alpha -}) P_{\bm{n}-\bm{\nu}_{\alpha -}}(t) \right. \nonumber \\
- T_{\alpha +}(\bm{n}&+&\bm{\nu}_{\alpha +}|\bm{n})
P_{\bm{n}}(t) - \left.
T_{\alpha -}(\bm{n}+\bm{\nu}_{\alpha -}|\bm{n})
P_{\bm{n}}(t) \right]\,,
\label{master_SLVC}
\end{eqnarray}
where $\bm{\nu}_{\alpha +}$ specifies the number of individuals of type $\alpha$ which increase during the reaction $\alpha\,+$ and $\bm{\nu}_{\alpha -}$ specifies the number of individuals of type $\alpha$ which decrease during the reaction $\alpha\,-$. So, $\bm{\nu}_{\alpha +} = (0,\ldots,0,1,0,\ldots,0)$ and $\bm{\nu}_{\alpha -} = (0,\ldots,0,-1,0,\ldots,0)$, the nonzero entries being in the $\alpha^{\rm th}$ position. Equations \eqref{trans_rates_SLVC} and \eqref{master_SLVC}, together with an initial condition for $P_{\bm{n}}$, completely specify the stochastic dynamics, so that we can, in principle, find $P_{\bm{n}}(t)$ for all $t$. 

We now make the diffusion approximation, mentioned above, that is, $V$ is assumed sufficiently large that $x_\alpha \equiv n_\alpha/V$ is approximately continuous~\cite{crow_kimura_into}. The other aspect of the approximation involves expanding out the master equation as a power series in $V^{-1}$, and neglecting powers of $V^{-3}$ and higher. The master equation for $P_{\bm{n}}(t)$ then becomes a Fokker-Planck equation (FPE) for $P(\bm{x}, t)$~\cite{gardiner_2009}:
\begin{eqnarray}
\frac{\partial P(\bm{x},t)}{\partial t} &=& 
- \frac{1}{V}\sum_{\alpha=1}^M \frac{\partial }{\partial x_\alpha} 
\left[ A_{\alpha}(\bm{x}) P(\bm{x},t) \right] \nonumber \\
&+& \frac{1}{2V} \sum_{\alpha,\beta=1}^M \frac{\partial^2 }{\partial x_\alpha \partial x_\beta} 
\left[ B_{\alpha \beta}(\bm{x}) P(\bm{x},t) \right].
\label{FPE_SLVC}
\end{eqnarray}
The precise form of the functions $A_\alpha(\bm{x})$ and $B_{\alpha \beta}(\bm{x})$ are found by carrying out the expansion, but explicit expressions for them also exist in terms of the $\bm{\nu}_{\alpha \pm}$ and the transition rates~\cite{mckane_BMB}. One finds that
\begin{eqnarray}
A_\alpha(\bm{x}) &=& \left( b_\alpha - d_\alpha \right)x_\alpha - \sum^{M}_{\beta = 1} c_{\alpha \beta } x_\alpha x_\beta \,, \nonumber \\
B_{\alpha \alpha}(\bm{x}) &=& \left( b_\alpha + d_\alpha \right)x_\alpha + \sum^{M}_{\beta = 1} c_{\alpha \beta } x_\alpha x_\beta \,,
\label{explicitA_B_SLVC}
\end{eqnarray}
and $B_{\alpha \beta}=0$, for $\alpha \neq \beta$. 

The FPE (\ref{FPE_SLVC}) is useful for systems with one degree of freedom. However for those with more than one degree of freedom, it is very difficult to analyse, and just as importantly, it is difficult to understand intuitively. For this reason we move over to the completely equivalent, but very different, formulation in terms of a set of stochastic differential equations (SDEs). For the FPE (\ref{FPE_SLVC}) these take the form~\cite{gardiner_2009}
\begin{equation}
\frac{\mathrm{d}x_\alpha}{\mathrm{d}\tau} = A_\alpha(\bm{x}) + \frac{1}{\sqrt{V}}\,\eta_\alpha(\tau), \ \ \alpha=1,\ldots,M,
\label{SDE_SLVC}
\end{equation}
where $\tau=t/V$, $\eta_\alpha(\tau)$ is a Gaussian white noise with zero mean and with a correlator 
\begin{equation}
\langle \eta_\alpha(\tau) \eta_\beta(\tau') \rangle = B_{\alpha \beta}(\bm{x}) \delta(\tau - \tau'),
\label{correlator}
\end{equation}
and where the SDE is to be interpreted in the sense of It\={o}. Equations (\ref{SDE_SLVC}) and (\ref{correlator}) together give the mesoscopic description of the system. The familiar, deterministic, Lotka-Volterra equations form the macroscopic description, and can be found by taking the $V \to \infty$ limit of Eq.~(\ref{SDE_SLVC}).

\subsection{The Moran model}
\label{sec:Moran_model_defn}
If we are to discuss the relationship of this model to the Moran model, we need to carry out a similar derivation to that given above, but for the Moran model, since we are not aware that the master equation or the Fokker-Planck equation appears in the literature for the case of $M$ alleles and selection, or at least not in the form that we require here. The derivation itself looks more complicated than the one for the SLVC model, due mainly to the fact that we have to implement the fixed $N$ constraint by expressing one variable in terms of the other $M-1$. Therefore we will only give the definition of the model in terms of the transition rates, and the final form for the FPE and the SDEs here, leaving the intermediate steps to Appendix~\ref{sec:Moran_setup}. 

The states of the system will be labelled by $n_1,\ldots,n_{M-1}$, since $n_M$ can be expressed in terms of the other $(M-1)$ through $n_M = N - \sum^{M-1}_{a = 1} n_a$. We will also use the notation $\underline{n} = (n_1,\ldots,n_{M-1})$. If we write $n_M$, then it should be understood as being equal to $n_M = N - \sum^{M-1}_{a = 1} n_a$. In what follows Greek indices $\alpha, \beta, \gamma, \ldots$ always run from $1$ to $M$ and Roman indices $a, b, c, \ldots$ always run from $1$ to $(M-1)$. 

\subsubsection{The neutral Moran model}

First of all, suppose there is no selection. This case is discussed in the literature; the transition rates are given by~\cite{blythe_mckane_models_2007}
\begin{equation}
T(n_1,\ldots,n_a + 1,\ldots,n_b - 1,\ldots,n_{M-1}|\underline{n}) = 
\frac{n_a}{N}\,\frac{n_b}{N},
\label{Moran_trans_1}
\end{equation}
with $a \neq b$, and
\begin{equation}
T(n_1,\ldots,n_a \pm 1,\ldots,n_{M-1}|\underline{n}) = 
\frac{n_a}{N}\,\frac{N - \sum^{M-1}_{b = 1} n_b}{N},
\label{Moran_trans_2}
\end{equation}
if either allele $a$ increases at the expense of allele $M$, or allele $M$ increases at the expense of allele $a$, respectively.

\subsubsection{The Moran model with frequency-independent selection}\label{sec_moran_FI_model}

We can now add constant selection, that is, selection for each allele that does not depend on population composition. Suppose that $W_\alpha$ is the fitness weighting of allele $\alpha$, $\alpha=1,\ldots,M$. Then Eqs.~(\ref{Moran_trans_1}) and (\ref{Moran_trans_2}) become
\begin{equation}
T(n_1,\ldots,n_a + 1,\ldots,n_b - 1,\ldots,n_{M-1}|\underline{n}) = 
\frac{W_a n_a}{\mathcal{W}}\,\frac{n_b}{N},
\label{Moran_trans_sel_1}
\end{equation}
if $a \neq b$;
\begin{equation}
T(n_1,\ldots,n_a + 1,\ldots,n_{M-1}|\underline{n}) = 
\frac{W_a n_a}{\mathcal{W}}\,\frac{N - \sum^{M-1}_{b = 1} n_b}{N},
\label{Moran_trans_sel_2}
\end{equation}
and
\begin{equation}
T(n_1,\ldots,n_a - 1,\ldots,n_{M-1}|\underline{n}) = \frac{n_a}{N}\,
\frac{W_M \left[ N - \sum^{M-1}_{b = 1} n_b \right]}{\mathcal{W}},
\label{Moran_trans_sel_3}
\end{equation}
where
\begin{equation}
\mathcal{W} = \sum^{M-1}_{a = 1} W_a n_a + W_M\left[ N - \sum^{M-1}_{b = 1} n_b \right].
\label{curly_W}
\end{equation}

We can simplify these expressions somewhat by taking the limit of weak selection. To do this we express $W_\alpha$ as equal to unity plus a small deviation of order $s$:
\begin{equation}
W_\alpha = 1 + s \rho_\alpha,
\label{selection_s}
\end{equation}
where $\rho_\alpha$ is of order one but can be positive or negative. However this still leads to rather cumbersome expressions and the details are given in Appendix~\ref{sec:Moran_setup}. There we also show that going over to new (continuous) variables $x_a = n_a/N$, one finds that the system can be described by the FPE
\begin{eqnarray}
\frac{\partial P}{\partial t} &=& - \frac{1}{N}\,\sum^{M-1}_{a = 1} \frac{\partial }{\partial x_a} \left[ A_a \left( \underline{x}\right) P\left( \underline{x},t\right) \right] \nonumber \\
&+& \frac{1}{2 N^2}\,\sum^{M-1}_{a,b=1} \frac{\partial^2 }{\partial x_a \partial x_b} \left[ B_{a b}\left( \underline{x}\right) P\left( \underline{x},t\right) \right],
\label{FPE_Moran}
\end{eqnarray}
with 
\begin{equation}
A_a(\underline{x}) = sx_a \left[ \rho_a - \sum^{M-1}_{b=1} \rho_b x_b - \rho_M \left( 1 - \sum^{M-1}_{b = 1} x_b \right) \right],
\label{Moran_A}
\end{equation}
to first order in $s$ and
\begin{equation}
B_{a b} (\underline{x}) = 2 \left( x_a\delta_{a b} - x_a x_b \right) + \mathcal{O}\left( s\right).
\label{Moran_B}
\end{equation}
As with our treatment of the SLVC model, we note that the above FPE is equivalent to an It\={o} SDE
\begin{equation}
\frac{\mathrm{d}x_a}{\mathrm{d}\tau} = A_a(\underline{x}) + \frac{1}{\sqrt{N}}\,\eta_a(\tau), \ \ a=1,\ldots,M-1\,,
\label{SDE_Moran}
\end{equation}
where $\tau=t/N$ and $\eta_a(\tau)$ is a Gaussian white noise with zero mean and with a correlator 
\begin{equation}
\langle \eta_a(\tau) \eta_b(\tau') \rangle = B_{ab}(\underline{x}) \delta(\tau - \tau').
\label{Moran_correlator}
\end{equation}
This is very similar to Eqs.~(\ref{SDE_SLVC}) and (\ref{correlator}), but with indices $a$ and $b$ replacing $\alpha$ and $\beta$ and $N$ replacing $V$, and with the functions $A_a$ and $B_{a b}$ taken from Eq.~(\ref{Moran_A}) and  Eq.~(\ref{Moran_B}).

The result for $B_{a b}(\underline{x})$ is just that of the neutral case, and has been known for a long time~\cite{kimura_1955}. The result for $A_a(\underline{x})$ can be checked by directly calculating $\mathrm{d}x_a/\mathrm{d}\tau$ from the master equation, as described in Appendix~\ref{sec:Moran_setup}. As we are explicitly considering the mapping between the models in the limit of weak selection, there is no need to go to higher order in $s$.

\subsubsection{The Moran model with frequency-dependent selection}
\label{sec_moran_FD_model}
We now consider the dynamics of the Moran model if the selective advantage experienced by an allele is dependent on the composition of the population. In this scenario the fitness of an allele $a$ is now denoted by $W_{a}(\underline{n})$, the inclusion of the explicit $\underline{n}$ argument indicating the dependence of the fitness on the nature of the population. The equations for the transition rates then take a similar form to those in the case when selection was constant, Eqs.~(\ref{Moran_trans_sel_1})-(\ref{curly_W}), but with $W_{a}$ replaced with $W_{a}(\underline{n})$. 

We are now left with a choice about how the fitness function $W_{a}(\underline{n})$ depends on the population composition. A common approach is to set $W_{a}(\underline{n})$ to a constant reproductive rate, moderated by a payoff from a game that each allele ``plays'' with every other allele in the population~\cite{nowak_2006}. There are many distinct ways to implement this, however, in line with \cite{antal_2009} we make the specific choice 
\begin{equation}
W_{\alpha}(\underline{n}) =  1 + s \left[ \sum_{b=1}^{M-1} g_{\alpha b} \frac{n_{b}}{N} + g_{\alpha M}\left( 1 - \sum_{b=1}^{M-1}\frac{n_b}{N} \right) \right] \,,
\label{W_freq_dep}
\end{equation}
where $g_{\alpha \beta}$ is the payoff to allele $\alpha$ from interacting with type $\beta$. 

As in the case of the Moran model with constant selection (addressed in Section~\ref{sec_moran_FI_model}), we can expand the master equation in terms of $1/N$ and $s$, and assuming that $N$ is large and $s$ small (formally $s\approx N^{-1}$), obtaining an approximation for the system dynamics in terms of an FPE of form Eq.~(\ref{FPE_Moran}). This is discussed in Appendix~\ref{sec:Moran_setup}, where it is shown that in this case $\underline{A}(\underline{x})$ is given by
\begin{eqnarray}
A_a(\underline{x}) &=& sx_a \left[ \mathcal{G}_{aM} + \sum^{M-1}_{b=1} G_{ab}x_b \right. \nonumber \\
&-& \left. \sum^{M-1}_{b=1} \mathcal{G}_{bM} x_b - \sum^{M-1}_{b,c=1} G_{bc} x_b x_c \right],
\label{Replicator_A}
\end{eqnarray}
to first order in $s$, while the form of $B(\underline{x})$ remains unchanged from that given in Eq.~(\ref{Moran_B}). Here we have defined the quantities
\begin{equation}
\mathcal{G}_{a \beta} \equiv g_{a \beta} - g_{M \beta}; \ \ 
G_{ab} \equiv \mathcal{G}_{ab} - \mathcal{G}_{aM}.
\label{relative_defns}
\end{equation}
Once again, this FPE is equivalent to an SDE of the form Eq.~(\ref{SDE_Moran}), but with $\underline{A}(\underline{x})$ taken from Eq.~(\ref{Replicator_A}).

It is interesting that it is the quantities $\mathcal{G}_{a M}$ and $G_{ab}$ which appear in the final expression for $A_a(\underline{x})$, and not simply $g_{a b}$. The quantity $\mathcal{G}_{a \beta}$ can be interpreted as a relative fitness, namely the payoff to allele $a$ against an opponent $\beta$ relative to the payoff to allele $M$ against the same opponent. Similarly, $G_{ab}$ is a relative relative fitness, namely the relative payoff to allele $a$ against an opponent with an allele $b$ relative to the relative payoff against an opponent with an allele $M$. 

At this order in $s$ the dynamics of the system in the deterministic $N\rightarrow \infty$ limit are equivalent to replicator dynamics~\cite{nowak_2004,hofbauer_1998}. As we discuss in the Introduction this equivalence does not hold at higher orders in $s$~\cite{traulsen_2005}. However, since we will work for the remainder of the paper in the weak selection limit, the mapping that we will develop between the SLVC model and the Moran model with frequency-dependent selection can also be interpreted as a mapping between the SLVC model and a stochastic version of the replicator dynamics.

\begin{figure}[t]
\setlength{\abovecaptionskip}{-2pt plus 3pt minus 2pt}
\begin{center}
\includegraphics[height=0.4\textwidth]{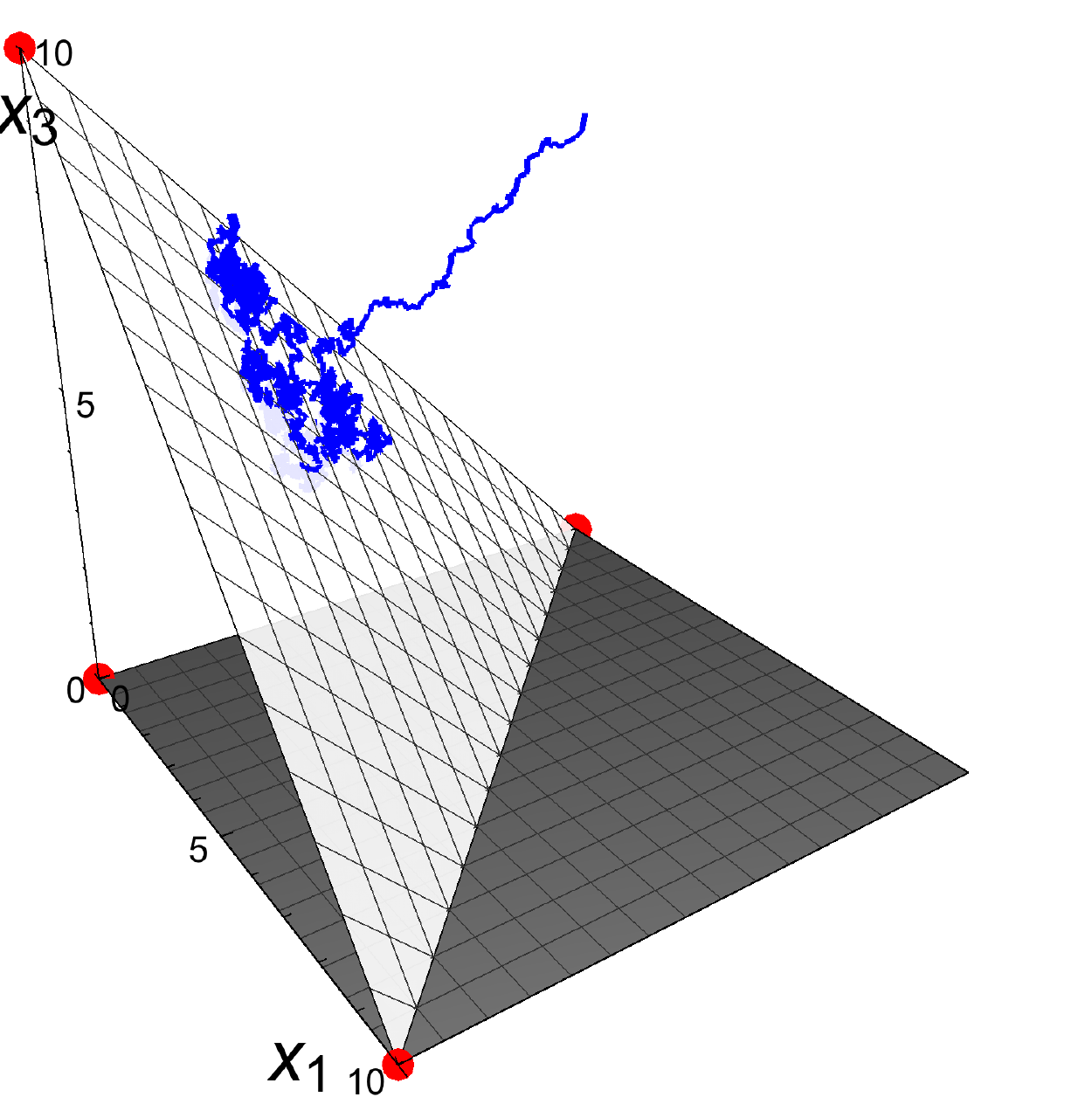}
\end{center}
\caption{(Color online) Plot illustrating a single realization of the stochastic dynamics (blue) of the neutral SLVC model (defined by Eqs.(\ref{trans_rates_SLVC}) and (\ref{master_SLVC})) with $\bm{n}$ scaled by $V$. Trajectories quickly collapse from $M$ to $M-1$ dimensions (here $M=3$ to $M=2$), after which the dynamics are constrained to the subspace specified by Eq.~(\ref{centre_manifold}), the deterministic center manifold (CM) (white plane). The system then moves neutrally within this plane until one of the absorbing states (red circles) is reached. Parameters used are given in Appendix~\ref{app_params}.}\label{fig_1}
\end{figure}

\section{Reduction of the Lotka-Volterra Model}
\label{sec:reduction}

In this section we will show that at medium to long-times the LV model with $M$ degrees of freedom, reduces to an $(M-1)$-dimensional model. We can then ask if there are any similarities between this reduced SLVC model and the Moran model. The reduction is accomplished by the systematic elimination of a fast mode, using techniques that we developed previously~\cite{constable_phys,constable_bio,constable_2015}. These require that we first understand the broad features of the deterministic ($V \to \infty$) dynamics, before we go on to study the stochastic dynamics. We begin with the neutral ($s=0$) model; effects due to selection will be introduced as perturbative corrections to the neutral case, since, as usual, we expect $s$ to be small. 

\subsection{The neutral model}
\label{sec:neutral}

The assumption that individuals of type $\alpha$, $\alpha=1,\ldots,M$, have equal fitness, that is, the theory is neutral, implies that they all have equal birth, death and competition rates: $b_\alpha \equiv b_0, d_\alpha \equiv d_0, c_{\alpha \beta} \equiv c_0$. In this case Eq.~(\ref{explicitA_B_SLVC}) reduces to 
\begin{eqnarray}
A_\alpha(\bm{x}) &=& x_\alpha \left[ \left( b_0 - d_0 \right) - c_0 \sum^{M}_{\beta = 1} x_\beta \right]\,, \nonumber \\
B_{\alpha \alpha}(\bm{x}) &=& x_\alpha \left[ \left( b_0 + d_0 \right) + c_0 \sum^{M}_{\beta = 1} x_\beta \right]\,.
\label{neutralA_B}
\end{eqnarray}
To characterize the deterministic dynamics, we first determine the fixed points of the dynamics by setting $A_\alpha(\bm{x})=0$ for all $\alpha$. It is clear that there are two classes of fixed points depending on whether $\sum^m_{\beta = 1} x_\beta$ is or is not equal to $(b_0 - d_0) c^{-1}_0$. If it is not, then $x_\alpha =0$ for all $\alpha$. So there is a fixed point at the origin and an $(M-1)$-dimensional hyperplane of fixed points given by
\begin{equation}
\sum^M_{\beta = 1} x_\beta = (b_0 - d_0)c_0^{-1}.
\label{centre_manifold}
\end{equation}

It is useful at this stage to rescale the $x_\alpha$ variables and time in order to eliminate the constants $b_0, c_0$ and $d_0$ from as much of the calculation as possible. To do so we introduce the new variables $y_{\alpha} = c_0 x_\alpha/(b_0 - d_0)$, $\alpha=1,\ldots,M$ and a new timescale $\tilde{\tau}= (b_0 - d_0)\tau$. Then the deterministic dynamics becomes 
\begin{equation}
\frac{\mathrm{d}y_\alpha}{\mathrm{d}\tilde{\tau}} = \tilde{A}_\alpha(\bm{y}) \equiv \frac{c_0}{(b_0 - d_0)^2}A_\alpha(\bm{x}) = y_\alpha \left[ 1 - \sum^M_{\beta = 1} y_\beta \right],
\label{rescaled_neutral_dyn}
\end{equation}
using Eq.~(\ref{neutralA_B}). The fixed points are now the origin and the $(M-1)$-dimensional hyperplane $\sum^M_{\beta = 1} y_\beta = 1$.  

Further insight can be gained by calculating the Jacobian
\begin{equation}
J_{\alpha \beta} = \frac{\partial \tilde{A}_\alpha}{\partial y_\beta} = \delta_{\alpha \beta} \left[ 1 - \sum^M_{\gamma = 1} y_\gamma \right] + y_\alpha \left[ - 1 \right],
\label{Jacobian}
\end{equation}
at points on this hyperplane. This is a highly degenerate matrix, with all columns identical to each other and equal to the matrix with entries $-y_\alpha$. It follows that there is a degenerate set of $(M-1)$ eigenvalues equal to zero, reflecting the existence of the hyperplane of fixed points. The remaining eigenvalue is equal to the trace of the Jacobian: $- \sum^M_{\beta =1} y_\beta = - 1$. We will label the zero eigenvalues $\lambda^{(a)}=0$, $a=1,\ldots,M-1$ and the non-zero eigenvalue $\lambda^{(M)}$. In addition, we will denote the left- and right-eigenvectors of the Jacobian corresponding to the eigenvalue $\lambda^{(\alpha)}$ by $\bm{u}^{(\alpha)}$ and $\bm{v}^{(\alpha)}$ respectively. They will be normalised so that $\sum^{M}_{\gamma=1}\,u^{(\alpha)}_\gamma v^{(\beta)}_\gamma = \delta_{\alpha \beta}$. For example, the left- and right-eigenvectors corresponding to the non-zero eigenvalue $\lambda^{(M)} = - 1$ are
\begin{align}\label{eigenvectors}
\bm{u}^{(M)} = \left( \begin{array}{c} 1 \\ \vdots \\ 1 \end{array} \right), 
\ \ \bm{v}^{(M)} = \left( \begin{array}{c} y_1 \\ \vdots \\ y_M \end{array} \right),
\end{align}
respectively. 

It is now possible to describe the neutral deterministic dynamics rather simply. If the system starts away from the hyperplane $\sum_\alpha y_\alpha = 1$, it will move towards the hyperplane at a rate governed by the non-zero eigenvalue of the Jacobian, that is, unity in the rescaled time and $(b_0 - d_0)^{-1}$ in the original time variable. Once it reaches the hyperplane, it remains at this point, since all points of the hyperplane are fixed points. The hyperplane is thus a center-manifold (CM) of the dynamics~\cite{wiggins_2003}. This dynamics is, of course, so simple that it is of limited interest, except that it forms the basis of the method that we will use to investigate the stochastic dynamics and of the dynamics with selection. For example, under the influence of weak noise, we would expect the system to similarly collapse onto the CM, but with a noisy trajectory, and once on the CM to move around purely stochastically. For this reason the direction perpendicular to the CM, $\bm{v}^{(M)}$, is called the \emph{fast} direction and the other directions, $\bm{v}^{(a)}$, $a=1,\ldots, (M-1)$, the \emph{slow} directions.

In our subsequent analysis we will need to map the initial condition of the SLVC model, $\bm{y}^{(0)}$, to an initial condition on the CM $\underline{y}^{\rm CM}$. This is most easily achieved by first introducing new variables to describe the deterministic dynamics of the neutral model. These are $f_a = x_a/\sum^M_{\beta = 1} x_\beta$, $a=1,\ldots,M-1$, and $\psi = \sum^M_{\beta = 1} x_\beta$. By direct substitution into the dynamical equations $dx_\alpha/dt = A_\alpha(\bm{x})$, with $A_\alpha(\bm{x})$ given by Eq.~(\ref{neutralA_B}), one finds that $df_a/dt = 0$, that is the $f_a$ are constants of the motion. This implies that they do not change from their initial values: $f_a = f^{(0)}_{a}$ at all times $t$. The only dyamical variable is $\psi$, which satisfies the equation $d\psi/dt = \psi [(b_0 - d_0) - c_0\psi]$. Therefore the dynamics simply consists of $\psi(t)$ decreasing (increasing) if it is larger (smaller) than $(b_0 - d_0)/c_0$, until it reaches this value. At this point the deterministic dynamics ceases, as the system is now on the CM. Using $f_a = f^{(0)}_{a}$ this point, labelled with the superscript CM, is given by
\begin{equation}
\frac{x^{\rm CM}_a}{\sum^M_{\beta = 1} x^{\rm CM}_{\beta}} = 
\frac{x^{(0)}_a}{\sum^M_{\beta = 1} x^{(0)}_{\beta}}.
\label{unchanging}
\end{equation}
Using $\sum^M_{\beta =1} x^{\rm CM}_\beta = (b_0 - d_0)/c_0$, and going over to the scaled variables $y_\alpha$, one finds that the initial condition on the CM is given by
\begin{equation}
y^{\rm CM}_a = \frac{y^{(0)}_a}{\sum^M_{\beta = 1} y^{(0)}_{\beta}}.
\label{IC}
\end{equation} 

A useful mathematical tool to separate the fast and slow dynamics is the projection operator defined by
\begin{equation}
P_{\alpha \beta} = \sum^{M-1}_{a = 1} v^{(a)}_{\alpha} u^{(a)}_{\beta}.
\label{projection_op}
\end{equation}
Suppose it is used to operate on the vector $\phi_\beta = \sum^M_{\kappa = 1} C_{\kappa} v^{(\kappa)}_\beta$, where the $C_{\kappa}$ are arbitrary constants. Then
\begin{eqnarray}
\sum^{M}_{\beta = 1} P_{\alpha \beta} \phi_\beta &=& \sum^{M}_{\beta,\kappa = 1} C_\kappa \sum^{M-1}_{a = 1} v^{(a)}_\alpha u^{(a)}_\beta v^{(\kappa)}_\beta 
\nonumber \\
&=& \sum^{M}_{\kappa = 1} C_\kappa \sum^{M-1}_{a = 1} v^{(a)}_\alpha \delta_{a \kappa} = \sum^{M-1}_{k = 1} C_k v^{(k)}_\alpha\,,
\label{proj_op}
\end{eqnarray}
that is, the fast term $C_M v^{(M)}_\beta$ has been wiped out.

We can use the projection operator directly as given by Eq.~(\ref{projection_op}), or observe that
\begin{equation}
P_{\alpha \beta} = \sum^{M}_{\gamma = 1} v^{(\gamma)}_{\alpha} u^{(\gamma)}_{\beta} -
v^{(M)}_{\alpha} u^{(M)}_{\beta} = \delta_{\alpha \beta} - v^{(M)}_{\alpha} u^{(M)}_{\beta},
\label{projection_operator}
\end{equation}
by completeness. Since $u^{(M)}_\beta = 1$ for all $\beta$, we may write this as
\begin{equation}
P_{\alpha \beta} = \delta_{\alpha \beta} - v^{(M)}_{\alpha} = \delta_{\alpha \beta} - y_\alpha.
\label{proj_operator}
\end{equation}
We also note that $\sum^M_{\alpha = 1} P_{\alpha \beta} = 0$.

The projection operator can be used to determine the stochastic dynamics \emph{on} the CM. In terms of the rescaled variables Eq.~(\ref{SDE_SLVC}) becomes 
\begin{equation}
\frac{\mathrm{d}y_\alpha}{\mathrm{d}\tilde{\tau}} = \frac{c_0}{(b_0 - d_0)^2}\frac{1}{\sqrt{V}}\,\eta_\alpha(\tau) \equiv \frac{1}{\sqrt{V}}\,\tilde{\eta}_\alpha(\tilde{\tau}), \ \ \alpha=1,\ldots,M,
\label{SDE_neut_scaled}
\end{equation}
since from Eq.~(\ref{neutralA_B}) we see that on the CM $A_\alpha(\bm{x})=0$. We also note that $B_{\alpha \beta} = 2b_0 x_\alpha\delta_{\alpha \beta}$ on the CM, and so
\begin{eqnarray}
& & \left\langle \tilde{\eta}_\alpha(\tilde{\tau}) \tilde{\eta}_\beta(\tilde{\tau}') \right\rangle = c^2_0 \left( b_0 - d_0 \right)^{-4}\,\left\langle \eta_\alpha(\tau) \eta_\beta(\tau') \right\rangle \nonumber \\
&=& c^2_0 \left( b_0 - d_0 \right)^{-4}\,\delta_{\alpha \beta}\,2 b_0 \frac{(b_0 - d_0)}{c_0}\,y_\alpha\,\left( b_0 - d_0 \right) \delta \left( \tilde{\tau} - \tilde{\tau}' \right) \nonumber \\
&=& 2 b_0 c_0 \left( b_0 - d_0 \right)^{-2}\,\delta_{\alpha \beta}\,y_\alpha\,\delta \left( \tilde{\tau} - \tilde{\tau}' \right).
\label{correlator_rescaled}
\end{eqnarray}
 
Application of the projection operator in Eq.~(\ref{proj_operator}) to $\mathrm{d}y_\beta/\mathrm{d}\tilde{\tau}$, which we denote as $\dot{y}_\beta$, gives $\dot{y}_\alpha - \sum^{M}_{\beta = 1} v^{(M)}_\alpha \dot{y}_\beta$. However on the CM, $\sum^{M}_{\beta = 1} \dot{y}_\beta = 0$, and we recover $\dot{y}_\alpha$. Therefore, defining a projected noise $\zeta_\alpha = \sum^{M}_{\beta = 1} P_{\alpha \beta} \tilde{\eta}_\beta$, the SDE (\ref{SDE_neut_scaled}) becomes
\begin{equation}
\frac{\mathrm{d}y_\alpha}{\mathrm{d}\tilde{\tau}} = \frac{1}{\sqrt{V}}\,\zeta_\alpha(\tilde{\tau}), \ \ \alpha=1,\ldots,M.
\label{SDE_proj_neut}
\end{equation}
Although, Eqs.~(\ref{SDE_neut_scaled}) and (\ref{SDE_proj_neut}) look similar, there are some significant differences. First, in Eq.~(\ref{SDE_proj_neut}), there are only $(M-1)$ independent variables (since $\sum_\alpha y_\alpha = 1$) and noises (since $\sum_\alpha P_{\alpha \beta} = 0$ implies that $\sum_\alpha \zeta_\alpha = 0$). Second, the noises, $\zeta_\alpha(\tilde{\tau})$, now have a different correlation function:
\begin{eqnarray}
\left\langle \zeta_\alpha(\tilde{\tau}) \zeta_\beta(\tilde{\tau}') \right\rangle &=& \sum^m_{\gamma = 1} \sum^m_{\kappa = 1} P_{\alpha \gamma} P_{\beta \kappa} \left\langle \tilde{\eta}_\gamma(\tilde{\tau}) \tilde{\eta}_\kappa(\tilde{\tau}') \right\rangle \nonumber \\ 
&=& \frac{2 b_0 c_0}{\left( b_0 - d_0 \right)^{2}}\,\delta \left( \tilde{\tau} - \tilde{\tau}' \right)\,\sum^m_{\gamma = 1} P_{\alpha \gamma} P_{\beta \gamma} y_\gamma. \nonumber \\
&=& \frac{2 b_0 c_0}{\left( b_0 - d_0 \right)^2}\,\left[ y_\alpha \delta_{\alpha \beta} - y_\alpha y_\beta \right]\,\delta \left( \tilde{\tau} - \tilde{\tau}' \right), \nonumber \\
\label{correlator_new}
\end{eqnarray}
using Eqs.~(\ref{proj_operator}) and (\ref{correlator_rescaled}).

In summary, the noisy dynamics on the CM is governed by the SDEs in Eq.~(\ref{SDE_proj_neut}), where the noise correlation is given by Eq.~(\ref{correlator_new}). However, as pointed out above, only $(M-1)$ of the $y_\alpha$ and the noises are independent. We therefore choose the dynamical variables to be the first $(M-1)$ $y_\alpha$ and denote these by $z_a$. 

The SDEs then become 
\begin{equation}
\frac{\mathrm{d}z_a}{\mathrm{d}\tilde{\tau}} = \frac{1}{\sqrt{V}}\,\zeta_a(\tilde{\tau}), \ \ a=1,\ldots,M-1,
\label{SDE_neut}
\end{equation}
where $\zeta_a(\tilde{\tau})$ is a Gaussian noise with zero mean and with a correlator
\begin{equation}
\left\langle \zeta_a(\tilde{\tau}) \zeta_b(\tilde{\tau}') \right\rangle 
= \frac{2 b_0 c_0}{\left( b_0 - d_0 \right)^2}\,\left[ z_a \delta_{a b} - z_a z_b \right]\,\delta \left( \tilde{\tau} - \tilde{\tau}' \right).
\label{correlator_in_z}
\end{equation}
The choice of independent variables is chosen to mirror those in the Moran model: the first $(M-1)$ being independent, with the final one being determined through the condition $y_M = 1 - \sum^{M-1}_{a=1} y_a$. In the Moran model this condition comes from the constraint $\sum^M_{\alpha=1} n_\alpha = N$, whereas in the SLVC model it comes from the equation of the CM.

\subsection{The model with selection}
\label{sec:selection}
Introducing selection into the neutral model implies that the birth, death and competition rates can now be different for different alleles. We write 
\begin{equation}
b_\alpha = b_0 \left( 1 + \epsilon \beta_\alpha \right), \ 
d_\alpha = d_0 \left( 1 + \epsilon \delta_\alpha \right), \ 
c_{\alpha \beta} = c_0 \left( 1 + \epsilon \gamma_{\alpha \beta} \right), 
\label{epsilon_intro}
\end{equation} 
where $\epsilon$ is a small selection constant, which will be taken to be proportional to the selection constant, $s$, in the Moran model, when we compare both models.

The SLVC model to first order in $\epsilon$ can now be constructed. There is no need to modify the $B_{\alpha \beta}(\bm{x})$ beyond their neutral form, as working in the weak selection limit, $\epsilon$ is small, and terms of order $\epsilon/N^{2}$ in the FPE can be neglected. The functions $A_\alpha(\bm{x})$ are given by (after rescaling as in Eq.(\ref{rescaled_neutral_dyn}))
\begin{eqnarray}
\tilde{A}_\alpha(\bm{y}) &=& \frac{c_0}{\left( b_0 - d_0 \right)^2} A_\alpha(\bm{x}) = y_\alpha + \epsilon \frac{\left( b_0 \beta_\alpha - d_0 \delta_\alpha \right)}{( b_0 - d_0 )} y_\alpha \nonumber \\
&-& y_\alpha \sum^M_{\beta = 1} y_\beta - \epsilon \sum^M_{\beta = 1} \gamma_{\alpha \beta} y_\alpha y_\beta + \mathcal{O}\left( \epsilon^2 \right).
\label{A_y_epsilon}
\end{eqnarray}
If we now search for fixed points of the dynamics: values of $\bm{y}$ such that $\tilde{A}_\alpha(\bm{y})=0$, it is found that generically there is at most one fixed point which is not on the boundary of the allowed region of $\bm{y}$ variables. This result is known to be true for the LV model~\cite{hofbauer_1998,noble_hastings_2011,gokhale_2010}; we infer the result for the reduced model by exploiting the mapping to the Moran model with frequency-dependent selection, which we will demonstrate in Sec.~\ref{sec:compare_freq_dep}. We therefore deduce that a CM does not exist for nonzero $\epsilon$.

Although a CM no longer exists, we would still expect the time-scale separation argument used previously for the neutral case to apply. The time-scale for the fast mode to collapse will only be changed by small terms of order $\epsilon$, but now the collapse will not be onto a CM, but onto an $(M-1)$-dimensional subspace on which there is a weak deterministic dynamics (of strength $\epsilon$) in addition to the same noise as was found in the neutral case. This subspace will not be planar, but we will nevertheless determine it by asking that there is no deterministic dynamics in the fast direction as defined by the eigenvector $\bm{v}^{(M)}$ found in the neutral case, i.e., we ask that $\bm{u}^{(M)}\cdot \tilde{\bm{A}}(\bm{y})=0$. An equivalent condition, using the explicit form for $\bm{u}^{(M)}$ given in Eq.~(\ref{eigenvectors}) is $\sum^M_{\alpha = 1} \tilde{A}_\alpha = 0$. This is an approximation, but in previous work where it has been used~\cite{constable_phys,constable_bio,constable_2015} it has been found to be a very good one. The condition is also consistent with the other terms which appear in the SDE, given in the neutral case by Eq.~(\ref{SDE_proj_neut}), that is, $\sum_\alpha \dot{y}_\alpha = 0$ and $\sum_\alpha \zeta_\alpha = 0$. In a similar fashion we will also continue to use Eq.~(\ref{IC}) as an approximation for the initial condition of the system on the slow subspace.

This condition on $\tilde{\bm{A}}(\bm{y})$ determines the equation of the slow manifold. To zeroth order in $\epsilon$ it gives $y_M = 1 -\sum^{M-1}_{a=1} y_a$, the equation of the CM. Therefore to determine the order $\epsilon$ correction to this equation which gives the slow manifold we write 
\begin{equation}
y_M = 1 - \sum^{M-1}_{a = 1} y_a + \epsilon f\left( y_1,\ldots,y_{M-1} \right) + \mathcal{O}(\epsilon^2) \label{slow_sub_eqn}
\end{equation}
where $f$ is a function to be determined. Substituting Eq.~(\ref{slow_sub_eqn}) into the condition $\sum^M_{\alpha = 1} \tilde{A}_\alpha = 0$ gives
\begin{eqnarray}
& & f\left( y_1,\ldots,y_{M-1} \right) = \sum^{M-1}_{a = 1} \frac{\left( b_0 \beta_a - d_0 \delta_a \right)}{(b_0 - d_0)}  y_a \nonumber \\
&+& \frac{\left( b_0 \beta_M - d_0 \delta_M \right)}{(b_0 - d_0)}\,\left( 1 - \sum^{M-1}_{a = 1} y_a \right) - \sum^{M-1}_{a,b = 1}\,\Gamma_{a b} y_a y_b \nonumber \\
&-& \sum^{M-1}_{a = 1}\,y_a \left\{ \gamma_{a M} + \gamma_{M a} - 2 \gamma_{M M} \right\} - \gamma_{M M},
\label{full_f}
\end{eqnarray}
where we have introduced the combinations of constants
\begin{equation}
\Gamma_{a b} = \gamma_{a b} - \gamma_{a M} - \gamma_{M b} + \gamma_{M M}.
\label{defn_Gamma}
\end{equation}
This is the same combination of $\gamma_{\alpha \beta}$ as appears in the definition of $G_{a b}$ given by Eq.~(\ref{relative_defns}), and the same interpretation in terms of relative quantities holds. In fact, using Eq.~(\ref{epsilon_intro}) we may introduce a similar quantity for the full competition rates $c_{\alpha \beta}$:
\begin{equation}
C_{a b} \equiv c_{a b} - c_{a M} - c_{M b} + c_{M M} = \epsilon c_0 \Gamma_{a b}.
\label{defn_C}
\end{equation}

We can now eliminate $y_M$ from the function $\tilde{A}_\alpha(\bm{y})$ given by Eq.~(\ref{A_y_epsilon}) to find
\begin{eqnarray}
& & \left.\tilde{A}_a(\underline{y})\right|_{\textrm{SS}} =  \epsilon y_a \left\{ \left[ \left( \Phi_a - \gamma_{a M} \right) - \left( \Phi_M - \gamma_{M M} \right) \right] \right. \nonumber \\
& & - \sum^{M-1}_{b=1} \left[ \left( \Phi_b - \gamma_{b M} \right) - \left( \Phi_M - \gamma_{M M} \right) \right] y_b \nonumber \\
& & \left. - \sum^{M-1}_{b = 1} \Gamma_{ab} y_b + \sum^{M-1}_{b,c = 1} \Gamma_{bc} y_b y_c \right\} + \mathcal{O}(\epsilon^2)
\label{A_tilde}
\end{eqnarray}
where SS indicates that this is $\tilde{A}_\alpha(\bm{y})$ evaluated on the slow-subspace and where, for clarity, we have introduced 
\begin{equation}
\Phi_a \equiv \frac{b_0 \beta_a - d_0 \delta_a}{b_0 - d_0}.
\label{Phi}
\end{equation}

Introducing the reduced variables $z_a$ as in the neutral case, we may add the term in Eq.~(\ref{A_tilde}) to the SDE given in Eq.~(\ref{SDE_neut}) to give the reduced SDE in the case with selection:
\begin{equation}
\frac{\mathrm{d}z_a}{\mathrm{d}\tilde{\tau}} = \left.\tilde{A}_a(\underline{z})\right|_{\textrm{SS}} + \frac{1}{\sqrt{V}}\,\zeta_a(\tilde{\tau}), \ \ a=1,\ldots,M-1,
\label{SDE_sel}
\end{equation}
where the noise is as in the neutral case, that is, with the correlator given by Eq.~(\ref{correlator_in_z}) and where $\left.\tilde{A}_a(\underline{z})\right|_{\textrm{SS}}$ is evaluated at first order in $\epsilon$.
 
In this section we have shown that even though the $M$-allele SLVC model begins with one more degree of freedom than the $M$-allele Moran model, after some time the extra degree of freedom decays away and the models begin to resemble each other. In the next section we seek to determine the precise conditions under which their dynamics are equivalent.

\begin{figure}[t]
\setlength{\abovecaptionskip}{-2pt plus 3pt minus 2pt}
\begin{center}
\includegraphics[height=0.25\textwidth]{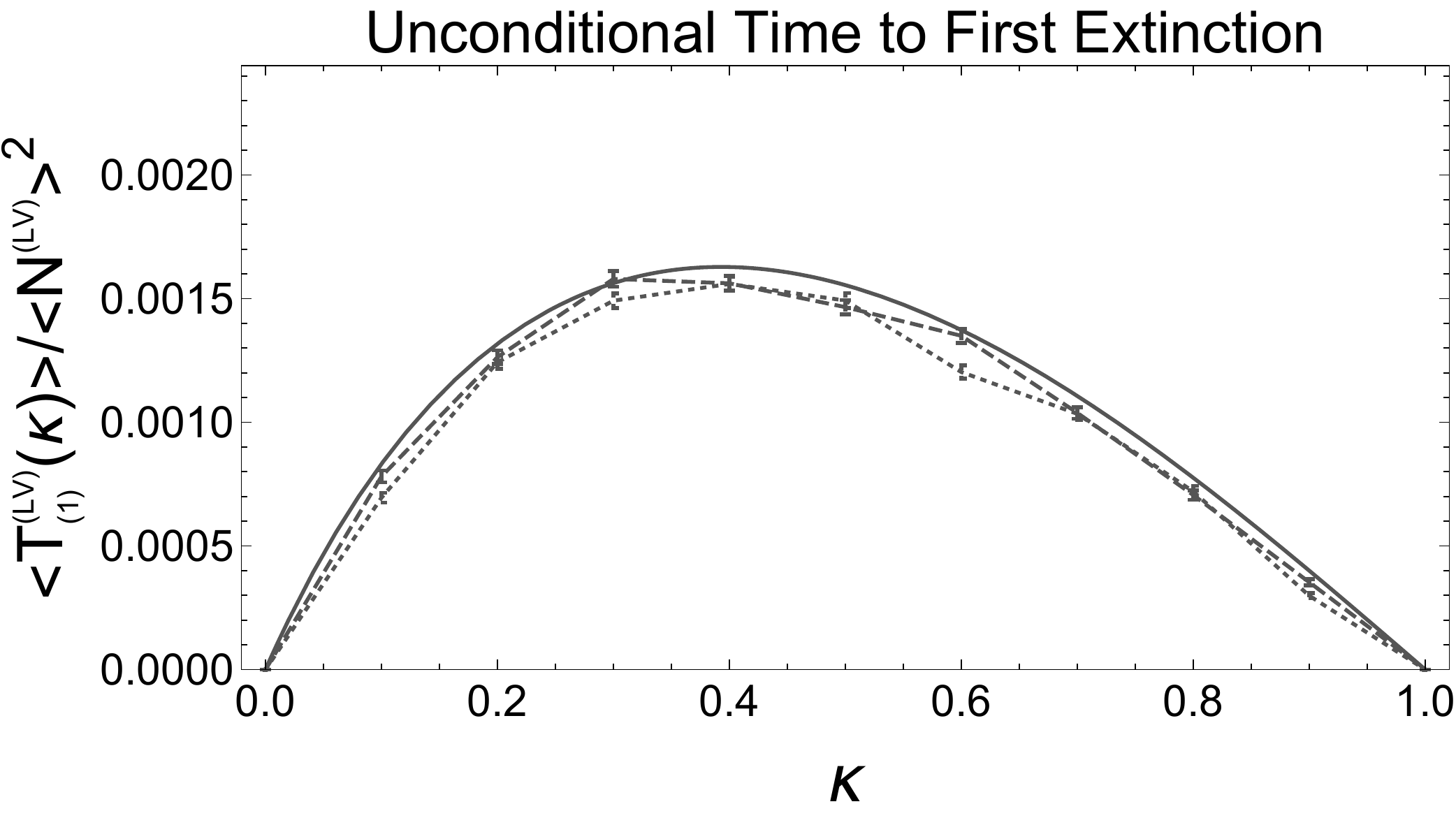}
\includegraphics[height=0.25\textwidth]{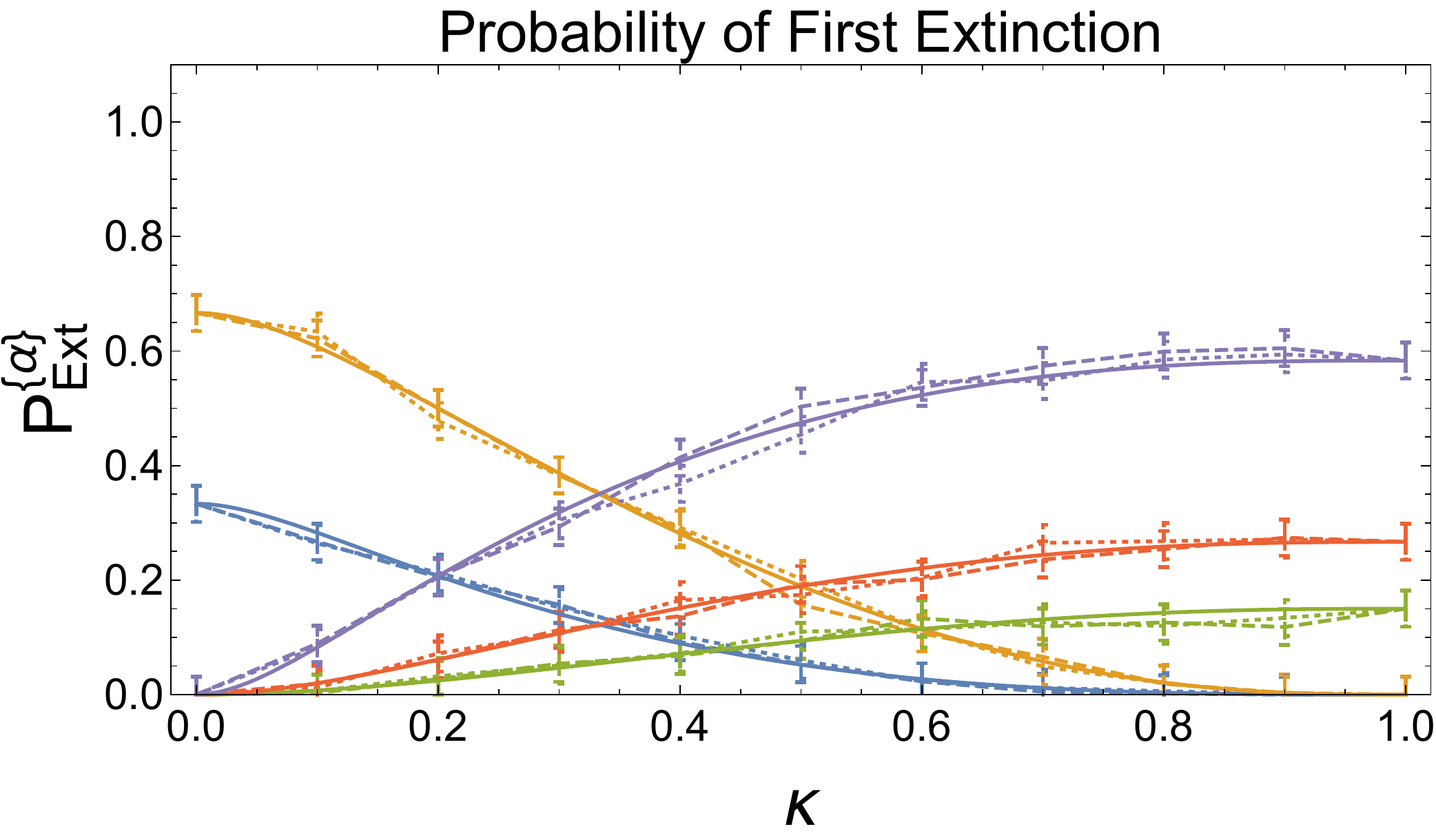}
\includegraphics[height=0.025\textwidth]{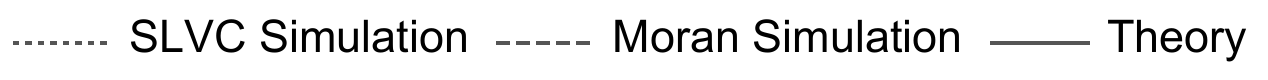}
\includegraphics[height=0.025\textwidth]{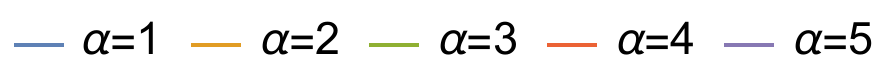}
\end{center}
\caption{(Color online) Plots of the time until the first extinction of an allele and the probability that each allele goes extinct first in the neutral models with $M=5$ species, plotted as a function of the projected initial condition on the center manifold. Simulation results are obtained from Gillespie simulation of Eqs.~(\ref{Moran_trans_1}) and (\ref{Moran_trans_2}) (Moran) and Eq.~(\ref{trans_rates_SLVC}) (SLVC), averaged over $10^{3}$ runs. Analytic results are obtained from Eq.~(\ref{eq_fixtime_SLVC_neutral}) with $r=1$ and the sum of Eq.~(\ref{eq_fixprob_SLVC_neutral}) over all extinction sequences that begin with a particular allele. In order to vary the initial condition on the CM as a function of a single variable, the initial condition for $\bm{x}^{(0)}$ in both the Moran and SLVC models has been parameterized by $\kappa$ (see Appendix~\ref{app_params}, where the remaining parameters are also given).}\label{fig_2}
\end{figure}

\section{Comparison of the reduced SLVC model and Moran models}
\label{sec:comparison}

\subsection{The neutral models}
\label{sec:compare_neutral}

It is clear that the reduced SLVC model written in the form of Eq.~(\ref{SDE_neut}) is precisely the neutral Moran model (Eqs.~(\ref{Moran_A}) and (\ref{Moran_B}) with $s=0$) up to a constant in the correlation function. The identification can be taken a little bit further by noting that the average population size of the SLVC model on the CM, Eq.~(\ref{centre_manifold}), in terms of the number of individuals in the system is 
\begin{eqnarray}
\langle N^{\mathrm{(LV)}} \rangle = (b_0 - d_0)V/c_0 \,. \label{eq_N_SLVC}
\end{eqnarray}
We can then transform Eq.~(\ref{SDE_neut}) back into a FPE equation in the natural units $t$ of the SLVC process, to obtain
\begin{eqnarray}
\frac{\partial P(\underline{z},t)}{\partial t} =  \frac{1}{2\langle N^{\mathrm{(LV)}} \rangle^{2}} \sum_{a,b=1}^{M-1} \frac{\partial^2 }{\partial z_a \partial z_b} 
\left[ \mathcal{B}_{ab}(\underline{z}) P(\underline{z},t) \right], \label{FPE_SLVC_Red_neutral}
\end{eqnarray}
where
\begin{equation}
\mathcal{B}_{ab}(\underline{z}) = 2 \frac{ b_0 }{ c_0 }(b_0 - d_0) \left( z_a\delta_{a b} - z_a z_b \right) \,. \label{SLVC_red_FPE_B}
\end{equation}
Comparing Eq.~(\ref{FPE_Moran}) (with  $\underline{A}(\underline{x})=0$ and $B(\underline{x})$ taken from Eq.~(\ref{Moran_B})) with Eq.~(\ref{FPE_SLVC_Red_neutral}), we see that in units of $t$, the natural timescale of the underlying stochastic processes, the timescale of the neutral SLVC model, is related to that of the Moran model by $t^{\mathrm{(M)}}=[b_0(b_0-d_0)/c_0]t^{\mathrm{(LV)}}$ for fixed $N=\langle N \rangle$. 

Some further algebra also allows us to calculate how the neutral Moran model and the neutral SLVC model are related in the SDE representation given Eqs.~(\ref{SDE_Moran}) and (\ref{SDE_SLVC}). Along with the scaling $x_a^{(\mathrm{M})}=c_0 x_a^{\mathrm{(LV)}}/(b_0-d_0)$, we find $\tau^{\mathrm{(M)}}=b_0\tau^{\mathrm{(LV)}}$. These results are summarized in Table~\ref{table_1}.

\
\begin{table*}[t]
    \begin{tabular}{| l | l | l | l | l |}
    \hline
  	 				&  All cases 									& Neutral 										& Constant selection											& Frequency-dependent selection \\ \hline
  Moran type 			& Eqs. (\ref{FPE_Moran}), (\ref{SDE_Moran})			& Eq. (\ref{Moran_B}), ($\underline{A}(\underline{x})=0$) 	& Eqs. (\ref{Moran_A}), (\ref{Moran_B}) 							& Eqs. (\ref{Replicator_A}), (\ref{Moran_B}) \\ 
  equations				& 											& 											& 								 						& \\ \hline
  SLVC  type 			& Eqs. (\ref{FPE_SLVC}), (\ref{SDE_SLVC}) 			& Eq. (\ref{explicitA_B_SLVC})						& Eq. (\ref{explicitA_B_SLVC})									& Eqs. (\ref{explicitA_B_SLVC})		  \\   
  equations				& 											& 											& 								 						& \\ \hline

  Mapping 				& $x_a^{\mathrm{(M)}}=c_0x_a^{\mathrm{(LV)}}/(b_0-d_0)$	& $\epsilon=s\equiv0$								& $\rho_\alpha = (b_0 \beta_\alpha - d_0 \delta_\alpha)/(b_0 - d_0) $ 		& $g_{\alpha M} = (b_0 \beta_\alpha - d_0 \delta_\alpha)/(b_0 - d_0)$\\ 
  					& $N^{\mathrm{(M)}}=(b_0-d_0)V/c_0$	 				& 	 										& $- \gamma_{aM}$											& $- \gamma_{aM}$ \\ 
  					& $\equiv\langle N^{(\mathrm{LV})}\rangle$				& 	 										& $s = \epsilon(b_0 - d_0)/b_0$									& $s = \epsilon(b_0 - d_0)/b_0$  \\ 
					& $t^{\mathrm{(M)}}=[b_0(b_0-d_0)/c_0]t^{\mathrm{(LV)}}$	&											& $\gamma_{ab} - \gamma_{aM} - \gamma_{Mb} + \gamma_{MM} = 0$		& $\gamma_{ab} - \gamma_{aM} - \gamma_{Mb}+\gamma_{MM} =$\\
					& $\tau^{\mathrm{(M)}}=b_0\tau^{\mathrm{(LV)}}$	 		& 	 										&														& $- \left[g_{ab} - g_{aM} - g_{Mb} + g_{MM}\right]$ \\ \hline
    \end{tabular}
\caption{Summary of the mappings between the Moran model in various forms and the SLVC model. Mappings are valid at long times in the limit where $N$ is large but finite, and $s$ and $\epsilon$ are small.}\label{table_1}
\end{table*}

\subsection{The models with frequency-independent selection}
\label{sec:compare_freq_indep}

In Sec.~\ref{sec:compare_neutral} we showed that the neutral reduced SLVC model and neutral Moran model are identical up to a rescaling of time. In this section we ask under what conditions the reduced SLVC model behaves identically to the Moran model with frequency-independent selection, defined by Eqs.~(\ref{FPE_Moran})-(\ref{Moran_B}).

This is carried out by comparing the constant terms, the terms linear in $y_b$ and the quadratic terms, in the bracket multiplying $\epsilon y_a$ in Eq.~(\ref{A_tilde}) with the corresponding bracket in Eq.~(\ref{Moran_A}). Comparing the constant terms gives $\rho_a - \rho_M = (\Phi_a - \gamma_{a M}) - (\Phi_M - \gamma_{M M})$. The linear terms now match if $\sum_b \Gamma_{ab} y_b = 0$ for all $y_b$, that is, if $\Gamma_{ab} = 0$ for all $a$ and $b$. This last condition implies that there is no quadratic term, as is clear from Eq.~(\ref{Moran_A}). Thus the condition for the mapping to exist is $\Gamma_{ab} = 0$ and the relationship between the $\rho_a$ and the $\gamma_{a M}$ that is required in order to carry out the mapping is $\rho_a = \Phi_a - \gamma_{a M} +$ constant. If we also identify $\rho_M = \Phi_M - \gamma_{M M}$, then this constant is zero, and we may then state that the reduced SLVC model behaves identically to the Moran model with frequency-independent selection if $\Gamma_{a b} \equiv 0$ and if we make the identification
\begin{equation}
\rho_\alpha = \frac{b_0 \beta_\alpha - d_0 \delta_\alpha}{b_0 - d_0} - \gamma_{\alpha M}.
\label{equiv_freq_indep}
\end{equation}

Assuming that this mapping can be carried out, we now transform Eq.~(\ref{A_tilde}) back into a FPE equation in the natural units $t$ of the SLVC process to obtain
\begin{eqnarray}
& &\frac{\partial P(\underline{z},t)}{\partial t} = 
- \frac{1}{\langle N^{\mathrm{(LV)}} \rangle}\sum_{a=1}^{M-1} \frac{\partial }{\partial z_a} 
\left[   \mathcal{A}_a\left( \underline{z} \right) P(\underline{z},t) \right] \nonumber \\
&+& \frac{1}{2\langle N^{\mathrm{(LV)}} \rangle^{2}} \sum_{a,b=1}^{M-1} \frac{\partial^2 }{\partial z_a \partial z_b} 
\left[   \mathcal{B}_{ab}\left( \underline{z} \right) P(\underline{z},t) \right],
\label{FPE_SLVC_Reduced_t}
\end{eqnarray}
where $\mathcal{A}_a(\underline{z})$ is a rescaled form of $\tilde{A}_a(\underline{y})|_{\textrm{SS}}$ --- reversing the rescaling carried out in Eq.~(\ref{A_y_epsilon}):
\begin{equation}
\mathcal{A}_a(\underline{z})= \frac{ (b_0 - d_0 )^2 }{c_0} \left. \tilde{A}_a(\underline{z})\right|_{\textrm{SS}},
\label{SLVC_red_FPE_A}
\end{equation}
and where $\mathcal{B}(\underline{z})$ retains the form given in Eq.~(\ref{SLVC_red_FPE_B}). Comparing Eqs.~(\ref{FPE_Moran})-(\ref{Moran_B}) with Eqs.~(\ref{FPE_SLVC_Reduced_t}), (\ref{SLVC_red_FPE_A}) and (\ref{SLVC_red_FPE_B}) we see that the two equations are identical if we rescale time in the SLVC such that $t^{\mathrm{(M)}}=[b_0(b_0-d_0)/c_0]t^{\mathrm{(LV)}}$ and simultaneously make the identification
\begin{equation}
s = \frac{\epsilon (b_0 - d_0)}{b_0},
\label{s_epsilon}
\end{equation}
along with the condition $\Gamma_{ab}=0$ for all $a$ and $b$. The equivalence of the reduced SLVC model and Moran model in the SDE setting is obtained by again enforcing the condition on $\Gamma_{ab}$ and (\ref{s_epsilon}), but rescaling time instead by $\tau^{\mathrm{(M)}}=b_0\tau^{\mathrm{(LV)}}$. We note that is different to the rescaling adopted in the analysis of the $M=2$ version of this correspondence~\cite{constable_2015}, where factors of $\gamma_{\alpha \beta}$ were included.

One can also ask about the nature of the fixed points in the Moran model and the reduced SLVC model under the conditions outlined above. From Eq.~(\ref{Moran_A}) the fixed points of the model are given by solutions to 
\begin{equation}
0 = x_a \left( \hat{\rho}_a - \sum^{M-1}_{b = 1} \hat{\rho}_b x_b \right),
\label{FP_Moran}
\end{equation}
where $\hat{\rho}_a \equiv \rho_a - \rho_M$. An analysis of this equation shows that the only fixed points are on a boundary, unless all the $\rho_{\alpha }$  are equal. However, if all the $\rho_{\alpha }$  are equal there is no selection, so in this case of frequency independent selection there are no interior fixed points.


\subsection{The models with frequency-dependent selection}
\label{sec:compare_freq_dep}

We can now repeat a similar analysis to that described in Sec.~\ref{sec:compare_freq_indep} to ask under what conditions the reduced SLVC model behaves identically to the Moran model with frequency-dependent selection, defined by Eqs.~(\ref{FPE_Moran}), (\ref{Replicator_A}) and (\ref{Moran_B}). 

Once again, we compare the constant terms, the terms linear in $y_b$ and the quadratic terms, in the bracket multiplying $\epsilon y_a$ in Eq.~(\ref{A_tilde}) with the corresponding bracket in Eq.~(\ref{Replicator_A}). Comparing the constant terms gives $g_{a M} - g_{M M} = (\Phi_a - \gamma_{a M}) - (\Phi_M - \gamma_{M M})$. The linear terms now match if $\sum_b G_{ab} y_b = - \sum_b \Gamma_{ab} y_b$ for all $y_b$, that is, if $G_{ab} = - \Gamma_{ab}$ for all $a$ and $b$. This last condition implies that the quadratic terms match. Thus the parameters of the two models are related by $\Gamma_{ab} = - G_{ab}$ and also by $g_{aM} = \Phi_a - \gamma_{a M} +$ constant. If we also identify $g_{MM} = \Phi_M - \gamma_{M M}$, then this constant is zero, and we may then state that the reduced SLVC model behaves identically to the Moran model with frequency-independent selection if $G_{ab} = - \Gamma_{a b}$ and if we make the identification
\begin{equation}
g_{\alpha M} = \frac{b_0 \beta_\alpha - d_0 \delta_\alpha}{b_0 - d_0} - \gamma_{\alpha M}.
\label{equiv_freq_dep}
\end{equation}
We may once again carry out rescalings and transformations as in Sec.~\ref{sec:compare_freq_indep} which lead to Eqs.~(\ref{FPE_SLVC_Reduced_t}), (\ref{SLVC_red_FPE_A}) and (\ref{s_epsilon}).

Under the conditions outlined above it can be shown that the deterministic frequency-dependent Moran model (equivalently the replicator equations) admits at most one stable fixed point on the interior region~\cite{gokhale_2010}. This is clearly also true for the reduced version of the SLVC model, since we have shown that it maps to the frequency-dependent Moran model. A more detailed discussion of these deterministic considerations can be found in~\cite{noble_hastings_2011}.

\begin{figure}
\setlength{\abovecaptionskip}{-2pt plus 3pt minus 2pt}
\begin{center}
\includegraphics[height=0.25\textwidth]{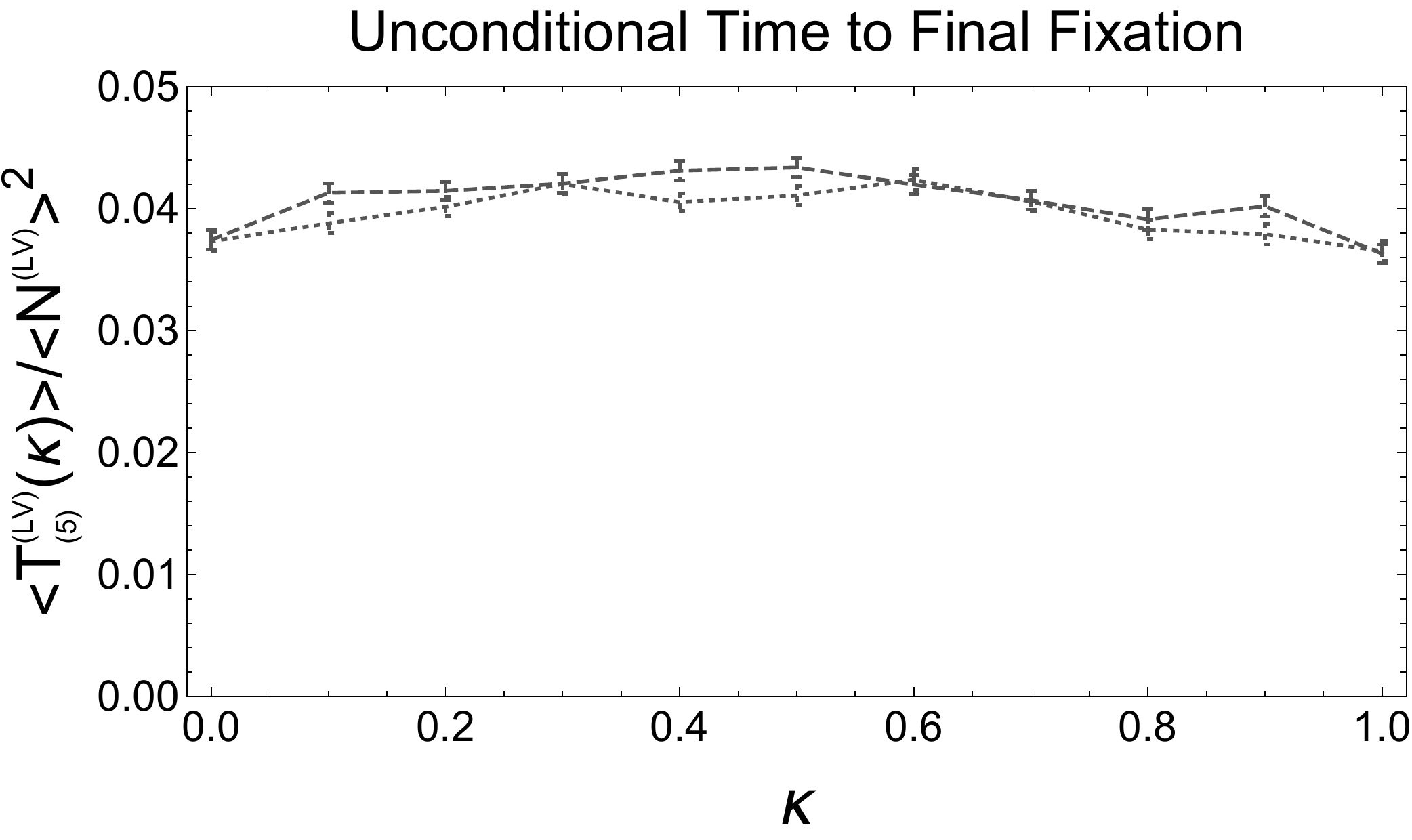}
\includegraphics[height=0.25\textwidth]{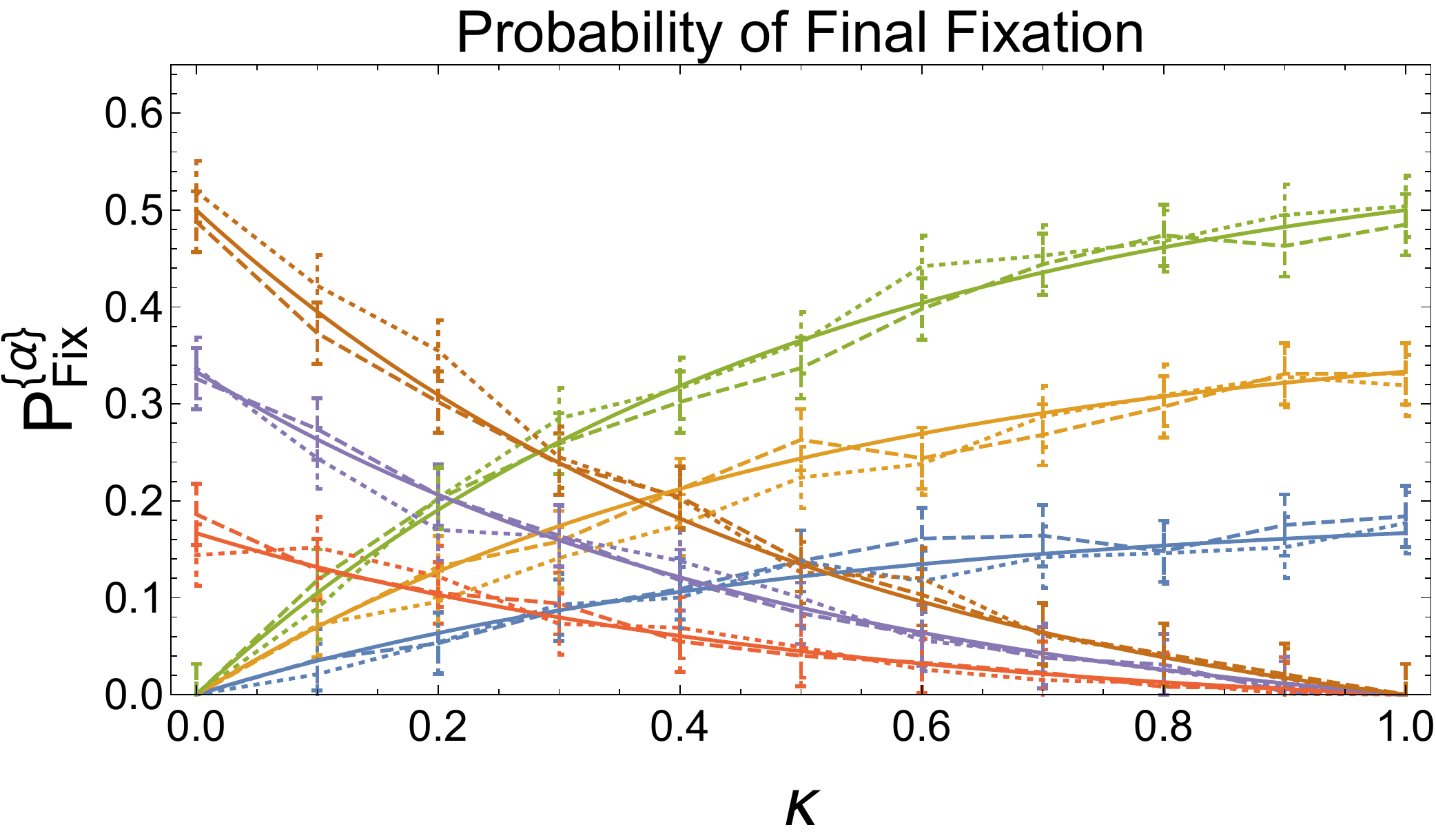}
\includegraphics[height=0.025\textwidth]{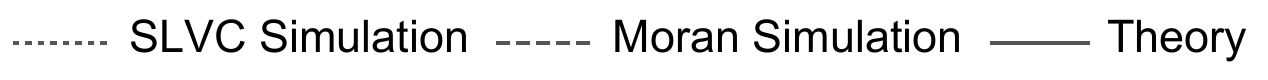}
\includegraphics[height=0.025\textwidth]{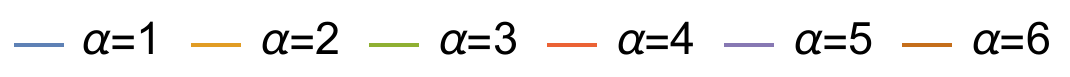}
\end{center}
\caption{(Color online) Plots of the unconditional mean time until the fixation of a single allele/species and the probability of the fixation of an allele/species for the Moran and SLVC models with frequency-independent selection in the case $M=5$ alleles/species. In these plots all alleles in the Moran model are under one of two selection pressures, while in the SLVC model all species have differing parameters that combine to give two selection pressures, making the system mappable to the Moran model presented. Analytic results for the probability of fixation are obtained using Eq.~(\ref{eq_fixprob_SLVC_indep}). Simulations results are the mean of $10^{3}$ stochastic simulations of the Moran and SLVC models. Parameters used are given in Appendix~\ref{app_params} where the parameterization of $\bm{x}^{(0)}$ in terms of $\kappa$ is also described.
}\label{fig_3}
\end{figure}

\section{Utilization of the Mappings}\label{sec_ins}

A useful feature of the mappings that have been derived above, is that we can now use analytic results obtained for the Moran model to make predictions about the SLVC model. In this section we illustrate this by mapping various results for fixation probabilities and times in the Moran model to those in the SLVC model, providing predictions for the extinction probabilities and mean time to extinction of species in certain scenarios.

\subsection{The neutral models}

In the neutral Moran system, it it possible to separate out the dynamics of the different alleles~\cite{baxter_2007}; rather than consider the dynamics of the many different interacting alleles, we can instead split the population into two subpopulations, one containing an allele of interest and the other containing all of the remaining alleles. Using this approach it is possible to calculate the probability of any series of extinctions and the time until the $r^{\rm th}$ extinction. These results, together with the mappings provided by Table~\ref{table_1}, allow us to calculate the same quantities for the SLVC model. The full calculation is detailed in Appendix~\ref{app_fixprob}, however here we state the main results for the SLVC model.

Let $\underline{S}=(\alpha_M, \alpha_{M-1},\ldots,\alpha_2)$ be a series of species extinctions in the neutral $M$-species SLVC model, such that species $\alpha_M$ goes extinct first, followed by $\alpha_{M-1}$ finally leaving only species $\alpha_1$ in the population. The probability of this series of extinctions, $P_{\mathrm{Fix}}^{\underline{S}}$, given initial conditions $\bm{x}^{(0)}$ is given by 
\begin{eqnarray}
P_{\mathrm{Fix}}^{\underline{S}} = \prod^{M-1}_{a=1} \frac{ x^{(0)}_{\alpha_a} }{ \sum^M_{\beta = 1} x^{(0)}_\beta - \sum_{b=1}^{a-1} x^{(0)}_{\alpha_b}  } \,.
\label{eq_fixprob_SLVC_neutral} 
\end{eqnarray}
We can also use this result to calculate the probability that species $\alpha$ goes extinct first, $P_{\mathrm{Ext}}^{\left\{ \alpha \right\}}$; this is simply the sum of Eq.~(\ref{eq_fixprob_SLVC_neutral}) over all $\underline{S}$ that do not contain $\alpha_{\alpha}$ as an element (i.e. species $\alpha$ never fixates but rather is the first to reach extinction). We find excellent agreement between this result and those obtained from simulations (see Figure~\ref{fig_2}).

We now move to considering the mean unconditional time until the extinction of the $r^{\rm th}$ species, $\langle T_{(r)}^{\mathrm{(LV)}}(\bm{x}^{(0)}) \rangle$. Given an initial distribution of species $\bm{x}^{(0)}$ in the SLVC model, this is given by 
\begin{eqnarray}
\langle T_{(r)}^{\mathrm{(LV)}}(\bm{x}^{(0)}) \rangle &=& - \frac{ c_0 \langle N^{(\mathrm{LV})} \rangle^{2} }{ b_0 ( b_0 - d_0 ) } \sum_{a=r}^{M-1} (-1)^{a-r} \binom{a-1}{r-1} \times \nonumber \\
      & & \sum_{ \bm{\alpha}} \left( \frac{ \sum_{b=1}^{a} x^{(0)}_{\alpha_{b}} }{ \sum_{\beta = 1}^{M} x^{(0)}_{\beta} }\right)\ln\left[ \frac{ \sum_{b=1}^{a} x^{(0)}_{\alpha_{b}} }{ \sum_{\beta=1}^{M} x^{(0)}_{\beta} } \right] \,, \nonumber \\ \label{eq_fixtime_SLVC_neutral}
\end{eqnarray}
where $ \langle N^{(\mathrm{LV})} \rangle$ is the average number of individuals in the SLVC model at carrying capacity (see Eq.~(\ref{eq_N_SLVC})). Here the summation over $\bm{\alpha}=\{ \alpha_1,\alpha_2,\ldots,\alpha_a\}$ denotes summation over all possible subsets of the set of positive integers $\{ 1,2,\ldots,M\}$ with $a$ entries. Again, this matches very well the results obtained from simulating the underlying stochastic models (see Figure~\ref{fig_2}). 

Note that in the Moran model, extinctions will occur on a time-scale proportional to $N^{2}$ when measured in units of $t^{\mathrm{(M)}}$ (see Appendix~\ref{app_fixprob}), while in the SLVC model we predict fixation to occur on a timescale $c_0 \langle N \rangle^{2} /\left[ b_0 ( b_0 - d_0 ) \right]$ (see Table \ref{table_1}). From this we see that increasing the birthrate, $b_0$, in the SLVC model increases the rate of `genetic drift' in the SLVC model relative to the Moran model. However a perhaps less intuitive result is that increasing the average genotype lifetime $b_0-d_0$ also increases the relative rate of genetic drift, while conversely increasing competition rate, $c_0$, slows down the rate of genetic drift in the SLVC model relative to the Moran model.

\subsection{The models with frequency-independent selection}

It is very difficult to obtain results on the fixation probability of an allele in a multi-allele Moran model when each of the alleles is under a different selection pressure, and to our knowledge no analytic results for this problem have been obtained. Progress can be made however if one assumes that all of the alleles are under one or other of two selective pressures. In this case the entire system dynamics can be decomposed into two processes; a Moran process with selection between the two subpopulations, and neutral drift within each population. The fixation probability of any allele can then be expressed as the product of the fixation probability of its subpopulation, multiplied by its fixation probability within the subpopulation. 

The full calculation is detailed in Appendix~\ref{app_fixprob}, while here we simply give the key results in the context of the SLVC model. Suppose that the $M$ species in the SLVC model are interacting in such a way that their dynamics can be described by frequency-independent selection (see Eq.~(\ref{equiv_freq_indep}) and Table~\ref{table_1}). Now further suppose that although their birth, death and interaction parameters may all be distinct, they are such that they only give rise to two distinct selection pressures when mapped to the Moran model with frequency-dependent selection; that is
\begin{eqnarray}
& &b_0\beta_\alpha - d_0 \delta_\alpha - (b_0 - d_0) \gamma_{\alpha M}= \nonumber \\ 
& &b_0\bar{\beta}_1 - d_0 \bar{\delta}_1 - (b_0 - d_0) \bar{\gamma}_{1 2} \quad 1 \leq \alpha \leq \theta \,, \nonumber \\ 
& &b_0\beta_\alpha - d_0 \delta_\alpha - (b_0 - d_0) \gamma_{\alpha M}=  \nonumber \\
& &b_0\bar{\beta}_2 - d_0 \bar{\delta}_2 - (b_0 - d_0) \bar{\gamma}_{2 2} \quad \theta < \alpha \leq M \,.
\label{parameter_map_1}
\end{eqnarray}
The fixation probability of any species in the SLVC model is then described by
\begin{eqnarray}
& &P_{\mathrm{Fix}}^{\{\alpha\} } = \nonumber \\
& &\frac{1 - \exp \left[ - \epsilon \langle N^{\mathrm{(LV)}} \rangle \omega_{12} \sum_{b=1}^{\theta} x^{(0)}_{b} /  \left( b_0\sum_{\beta=1}^{M} x^{(0)}_{\beta} \right) \right]}{1 -\exp \left[ - \epsilon \langle N \rangle \omega_{12} / b_0\right]} \times \nonumber \\
& &\frac{ x^{(0)}_{\alpha} }{ \sum_{\beta=1}^{\theta} x^{(0)}_{\beta}  } \, , \qquad \qquad \qquad \qquad \qquad 1 \leq \alpha \leq \theta \,,\nonumber \\
& &P_{\mathrm{Fix}}^{\{\alpha\} } = \nonumber \\
& &\left\{ 1 -  \frac{1 - \exp \left[ - \epsilon \langle N^{\mathrm{(LV)}} \rangle \omega_{12} \sum_{b=1}^{\theta} x^{(0)}_{b} /  \left( b_0\sum_{\beta=1}^{M} x^{(0)}_{\beta} \right) \right]}{1 -\exp \left[ - \epsilon \langle N \rangle \omega_{12} / b_0\right]} \right\} \times \nonumber \\
& &\frac{ x^{(0)}_{\alpha} }{\sum_{\beta = \theta + 1}^{M} x^{(0)}_{\beta} } \, , \qquad \quad \qquad \qquad \qquad \theta < \alpha \leq M \,, \nonumber \\
\label{eq_fixprob_SLVC_indep}
\end{eqnarray}
where
\begin{eqnarray}
\omega_{12} &=& (b_0(\bar{\beta_1} - \bar{\beta_2}) - d_0(\bar{\delta_1} - \bar{\delta_2}) - \nonumber \\
		& &(b_0-d_0)(\bar{\gamma}_{1 2} - \bar{\gamma}_{2 2}) ) \,,
\label{omega_12}
\end{eqnarray}
and where we have made use of Eqs.~(\ref{IC}) and (\ref{s_epsilon}). We find excellent agreement between these results and results obtained from simulations, as illustrated in Figure~\ref{fig_3}. Determining the probability of first extinction is not possible however; this requires knowing the probability of time ordering of extinctions (whether fixation in the subpopulation occurs before fixation in the population as a whole). 

More generally, we find good agreement between the Moran model and the SLVC model with conditions and mappings taken from Table~\ref{table_1}, even when analytic results are not available (see Figure~\ref{fig_5}).

\begin{figure}[t]
\setlength{\abovecaptionskip}{-2pt plus 3pt minus 2pt}
\begin{center}
\includegraphics[height=0.25\textwidth]{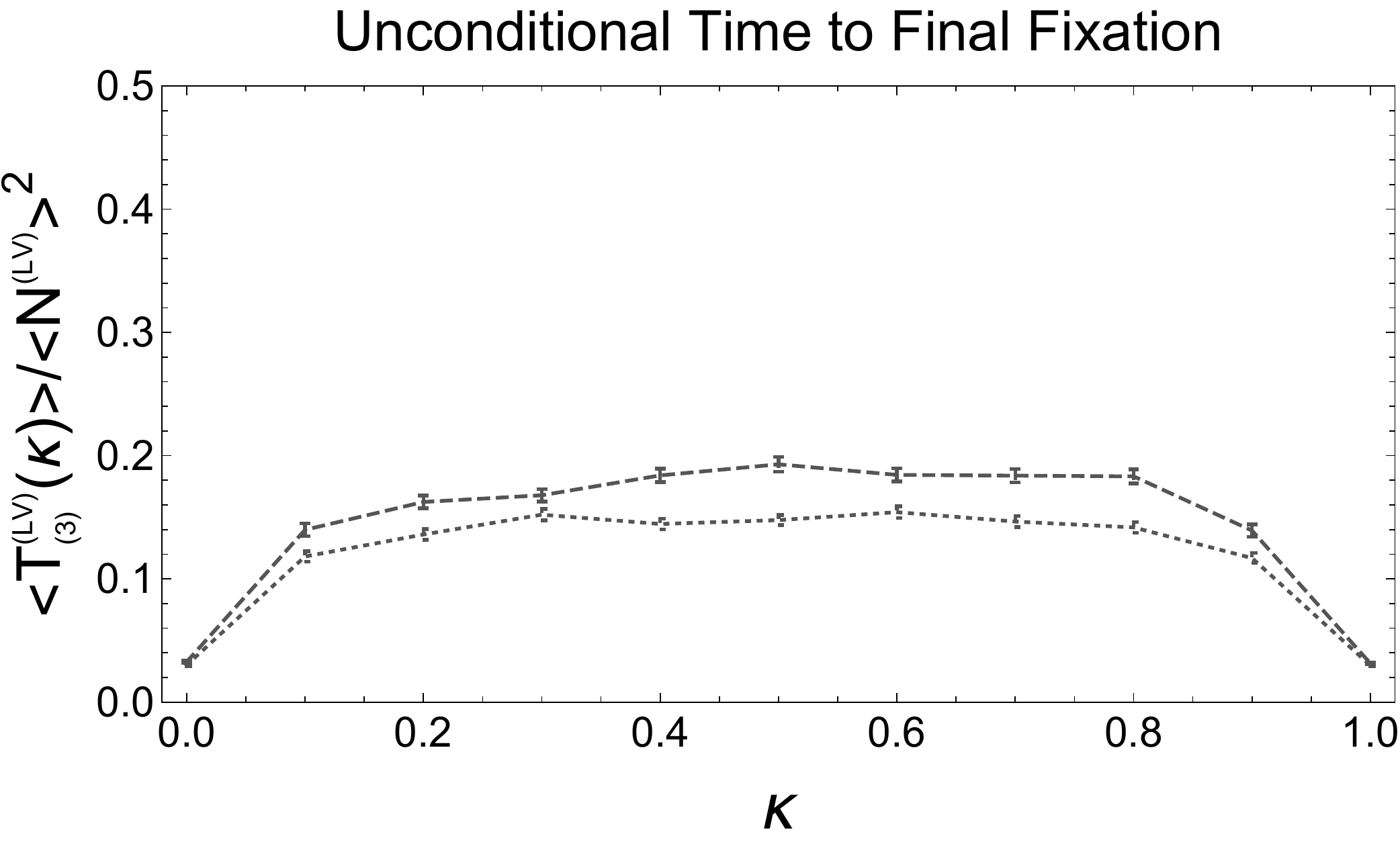}
\includegraphics[height=0.25\textwidth]{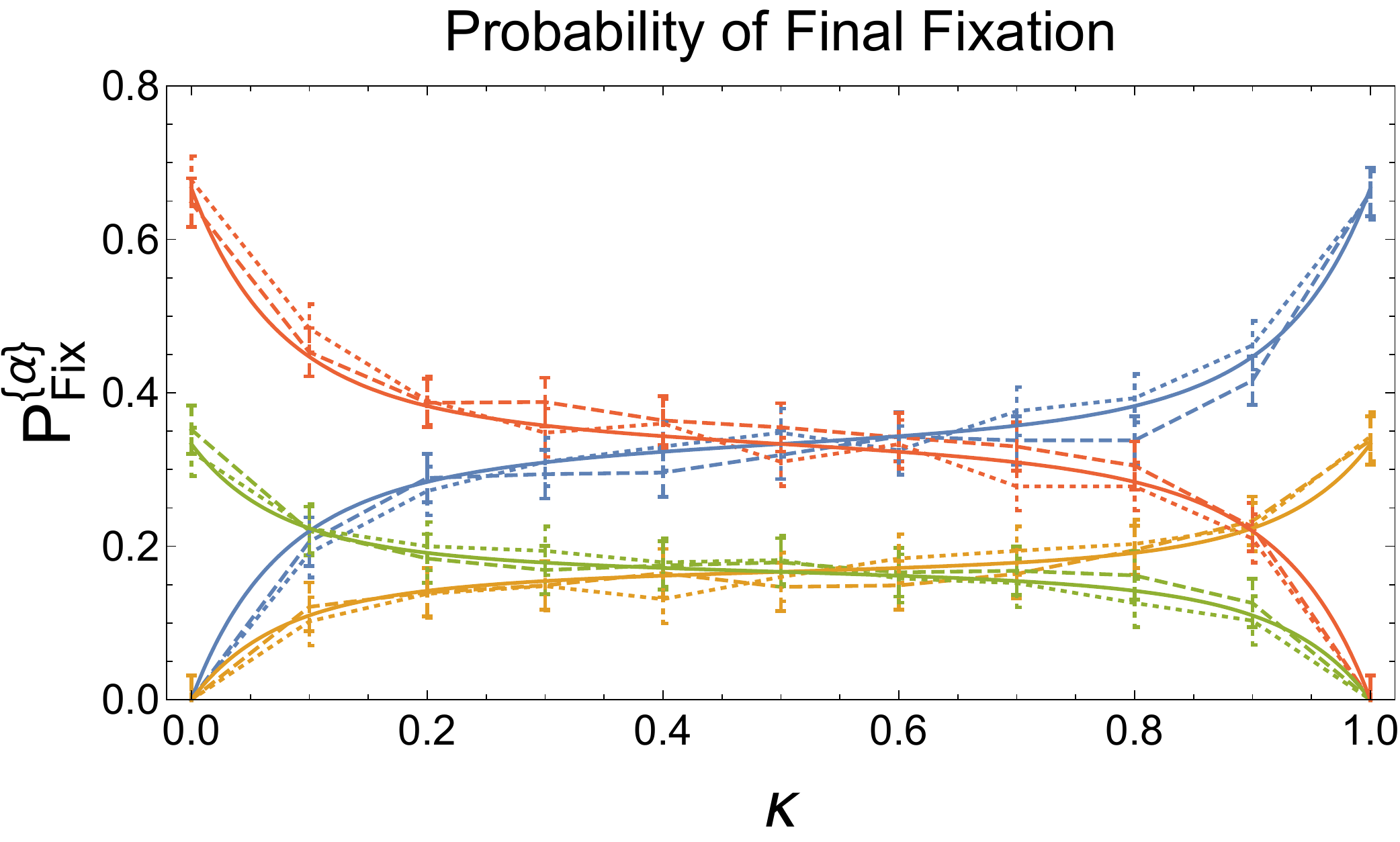}
\includegraphics[height=0.025\textwidth]{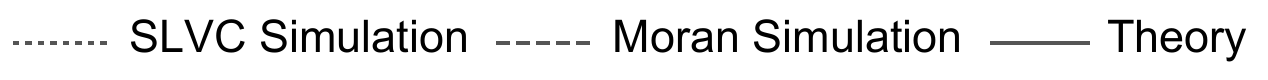}
\includegraphics[height=0.025\textwidth]{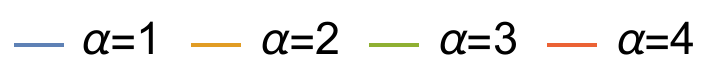}
\end{center}
\caption{(Color online) Plots of the unconditional mean time until the fixation of a single strategy/species and the probability of the fixation of a strategy/species for the Moran and SLVC models with frequency-dependent selection in the case of $M=4$ strategies/species. In these plots all players play one of two pairwise strategies, while in the SLVC model all species have one of two competition matrices, making the system mappable to the stochastic replicator model presented. Analytic results for the probability of fixation are obtained using Eq.~(\ref{eq_fixprob_SLVC_dep}). Simulations results are the mean of $10^{3}$ stochastic simulations of the Moran and SLVC models. Parameters used are given in Appendix~\ref{app_params} (where the parameterization of $\bm{x}^{(0)}$ in terms of $\kappa$ is also described), however we note that they are such that an interior fixed point exists for the deterministic dynamics. }\label{fig_4}
\end{figure}

\subsection{The models with frequency-dependent selection}\label{sec_ins_FD}

The inclusion of cubic terms in the multiallelic model model with frequency-dependent selection makes obtaining analytic results for the fixation probability and time even more challenging than in the case with frequency-independent selection. However analytic progress can again be made under the condition that the elements of the payoff matrix can be partitioned such that only two distinct strategies exists within the population (see Appendix~\ref{app_fixprob}). Here we describe the results of this calculation in relation to the SLVC model.

We first make the assumption that the competition matrix $\gamma_{ \alpha \beta }$ can be partitioned such that 
\begin{eqnarray}
\gamma_{\alpha \beta} &= \bar{\gamma}_{11} \,,\quad \alpha,\beta \leq \theta \,; \quad \gamma_{\alpha \beta} = \bar{\gamma}_{12} \,, \quad \alpha \leq \theta < \beta \,; \nonumber \\
\gamma_{\alpha \beta} &= \bar{\gamma}_{22} \,, \quad \alpha, \beta > \theta \,; \quad \gamma_{\alpha \beta} = \bar{\gamma}_{21} \,, \quad \beta \leq \theta < \alpha \nonumber \,,
\end{eqnarray}
while the birth and death terms can be partitioned such that\begin{eqnarray}
\beta_{\alpha} &= \bar{\beta}_{1} \,, \quad \delta_{\alpha} = \bar{\delta}_{1} \,, \quad 1 \leq \alpha \leq \theta \,, \nonumber \\
\beta_{\alpha} &= \bar{\beta}_{2} \,, \quad \delta_{\alpha} = \bar{\delta}_{2} \,, \quad \theta < \alpha \leq M \nonumber \,.
\end{eqnarray}
Then the fixation probability of any species in the population is shown in Appendix~\ref{app_fixprob} to be given by
\begin{eqnarray}
P_{\mathrm{Fix}}^{\{\alpha\} } &=& \frac{ 1 - \chi\left[ l(\bm{x}^{(0)}) \right]  }{ 1 - \chi\left[ l(1) \right] } \frac{ x^{(0)}_{\alpha} }{ \sum_{\beta=1}^{\theta} x^{(0)}_{\beta}  } \,, \quad \rm{if} \quad 1 \leq \alpha \leq \theta\,,\nonumber \\
P_{\mathrm{Fix}}^{\{\alpha\} } &=& \frac{ 1 - \chi\left[ l(\bm{x}^{(0)}) \right]  }{ 1 - \chi\left[ l(1) \right] } \frac{ x^{(0)}_{\alpha} }{ \sum_{\beta=\theta + 1}^{M} x^{(0)}_{\beta}} \,, \quad {\rm{if}} \quad \theta < \alpha \leq M \,,\nonumber \\
\label{eq_fixprob_SLVC_dep}
\end{eqnarray}
where
\begin{eqnarray}
\chi[l(\bm{x}^{(0)})] = \frac{\mathrm{erfi}\left[ l(\bm{x}^{(0)}) \right] }{\mathrm{erfi}\left[ l(0) \right] } \,, \quad \rm{if} \quad \bar{\Gamma}_{11} < 0 \,,\nonumber \\
\chi[l(\bm{x}^{(0)})] = \frac{\mathrm{erfc}\left[ l(\bm{x}^{(0)}) \right]  }{\mathrm{erfc}\left[ l(0) \right] } \,, \quad \rm{if} \quad \bar{\Gamma}_{11} > 0 \,. \label{eq_chi_M_x}
\end{eqnarray}
Here erfi and erfc are respectively the imaginary and complimentary error functions~\cite{handbook_1965,HTFS_1953}, and the function $l(\bm{x}^{(0)})$ is defined as
\begin{eqnarray}
l(\bm{x}^{(0)}) &=& \sqrt{ \frac{ \epsilon ( b_0 - d_0 ) \langle N^{\mathrm{(LV)}} \rangle }{2 b_0 |\bar{\Gamma}_{11}| } } \left( - \bar{\Gamma}_{11} \frac{ \sum_{b=1}^{\theta} x^{(0)}_{b} }{ \sum_{\beta=1}^{M} x^{(0)}_{\beta} }  \right. \nonumber \\
&+& \left. \frac{ b_0 ( \bar{\beta}_1 - \bar{\beta}_2 ) - d_0( \bar{\delta}_{1} - \bar{\delta}_{2} )}{b_0-d_0} - (\bar{\gamma}_{12} - \bar{\gamma}_{22}) \right) \,.\nonumber \\
\label{eq_l_x}
\end{eqnarray}
In addition, the notation $l(1)$ and $l(0)$ means the value of $l$ when $\left( \sum_{b=1}^{\theta} x^{(0)}_{b} \right) / \left(\sum_{\beta=1}^{M} x^{(0)}_{\beta} \right)  =1$ and $\left(\sum_{b=1}^{\theta} x^{(0)}_{b} \right) / \left(\sum_{\beta=1}^{M} x^{(0)}_{\beta} \right) = 0$, respectively. Using this analytic result we find good agreement between our theory and results obtained from simulations, as illustrated in Figure~\ref{fig_4}.

We note however that in Figure~\ref{fig_4} there begins to be disagreement for the time to fixation. This difference is likely caused by the existence of an interior stable fixed point in the system. In general, our method of fast-variable elimination is expected to perform well when $s$ is small and $N$ is relatively large~\cite{constable_phys}. However, when an interior fixed point is present, another factor comes into play; the validity of the diffusion approximation itself. As the distribution of stochastic trajectories about the fixed point becomes increasingly stable (for instance, by increasing $N$), the FPE obtained via the diffusion approximation is known to become less accurate near the boundaries~\cite{doering_2006}. Therefore, although our approximation of the FPE becomes more accurate with increasing $N$, the FPE itself becomes less reliable in this regime for increasing $N$. The disagreement hinted at here between the Moran-replicator and SLVC models will then only be calculable outside the diffusion limit. 

\begin{figure}
\setlength{\abovecaptionskip}{-2pt plus 3pt minus 2pt}
\begin{center}
\includegraphics[height=0.25\textwidth]{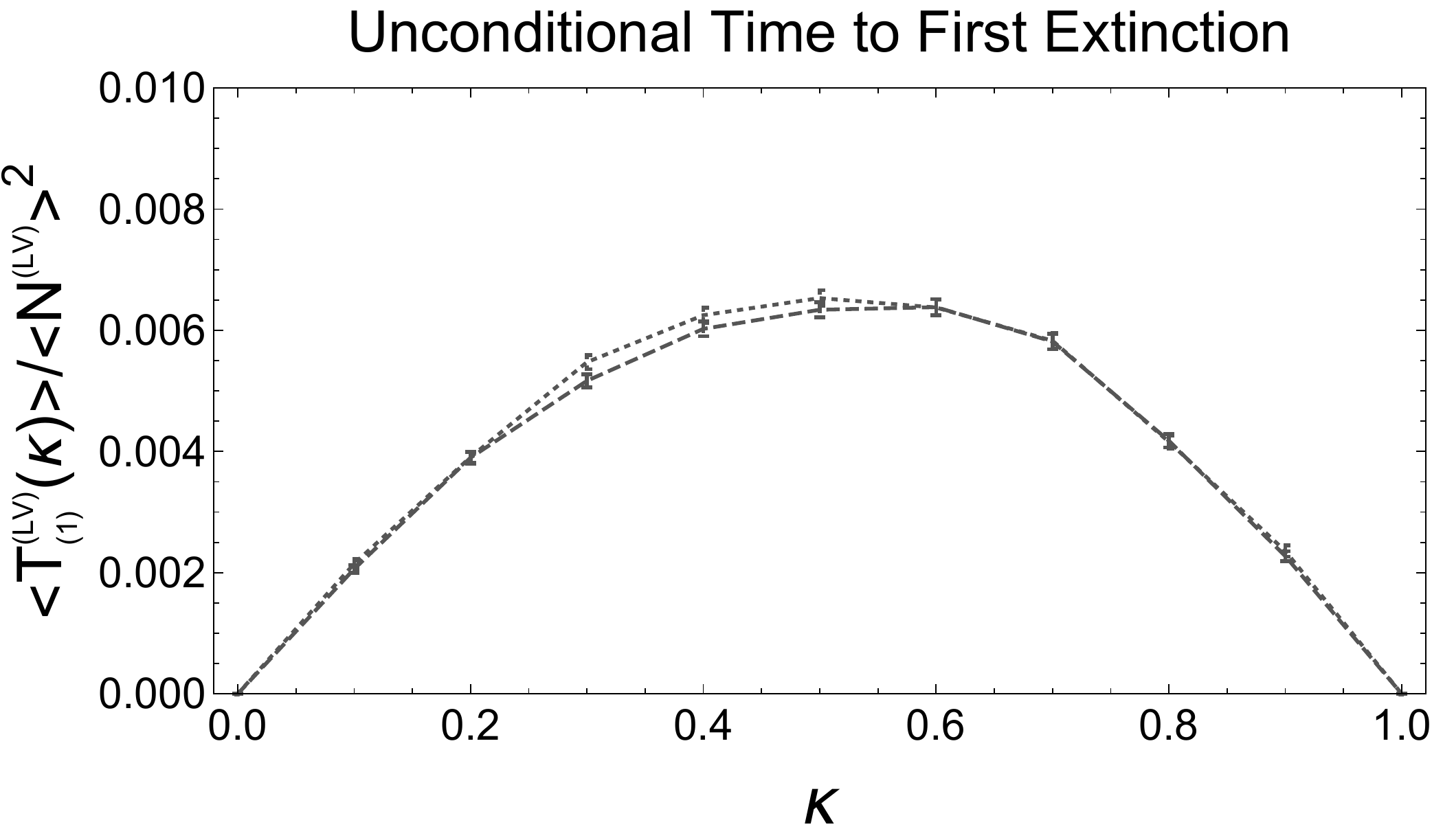}
\includegraphics[height=0.25\textwidth]{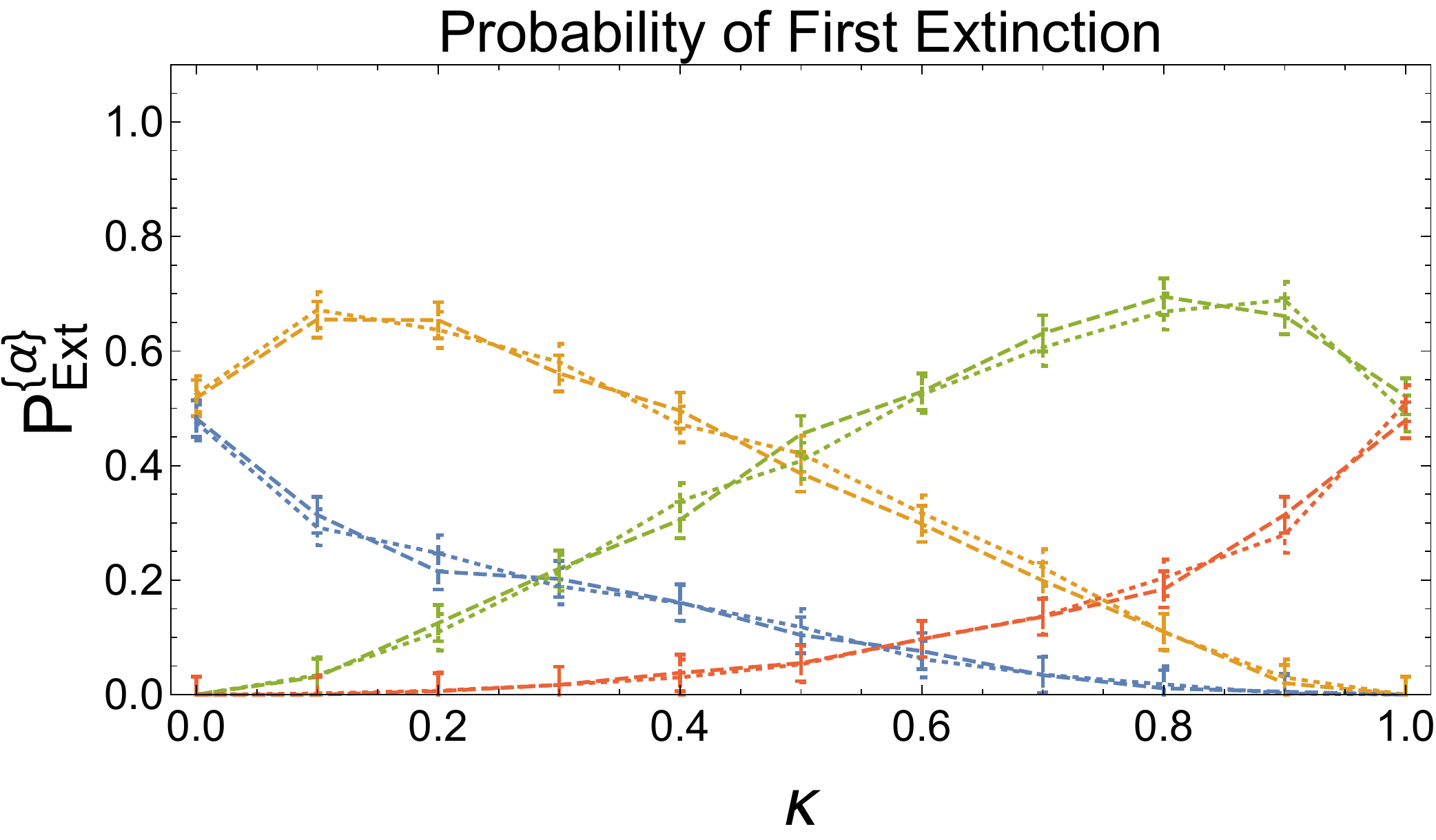}
\includegraphics[height=0.025\textwidth]{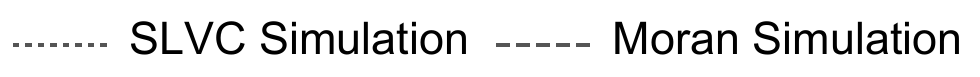}
\includegraphics[height=0.025\textwidth]{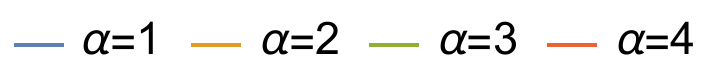}
\end{center}
\caption{
(Color online) Plots of the mean time until first extinction of an allele and the probability that each allele becomes extinct first in the Moran and SLVC models with frequency-dependent selection in the case of $M=4$ strategies/species. Each allele is under distinct selection pressures with differing game payoffs, $g$ or competition terms, $\gamma$. Full details of the parameters used are given in Appendix~\ref{app_params}. Although no analytic results for this system are available, we see good agreement between the Moran model and the SLVC model. Simulations are the average of $10^{3}$ runs.
}\label{fig_5}
\end{figure}

Solving for fixation times in this scenario becomes increasingly difficult; as we have discussed they are not straightforward to obtain even in a two-type system, as the emergence of very stable fixed points creates deviations between the underlying stochastic process and the diffusion approximation that have to be carefully corrected for~\cite{doering_2006}. Adding more types compounds this difficulty. However, comparing results of simulations of the Moran model with the SLVC model, we can see that the mapping between the models remains qualitatively intact (see Figure.~\ref{fig_4}) and that the mapping also holds under a broad range of parameter regimes in which there is no interior fixed point (see Figure \ref{fig_5}).

\section{Conclusions}

In this paper we have defined a mapping between various incarnations of the Moran model in $M-1$ variables and the SLVC model in $M$ variables, which accounts for demographic noise. In particular we have concentrated on the mapping between three forms of the Moran model: the neutral Moran model, the Moran model with frequency-independent selection and the Moran model with frequency-dependent selection. While the Moran model is formulated in terms of a population of fixed size $N$, the populations in the SLVC model do not strictly have a fixed size. However, at long times they approach a carrying capacity around which they stochastically fluctuate. The key to making an analytic bridge between these models has been in noting the following: if selection is weak and the population size is large, the SLVC model approaches its carrying capacity (and is confined in its vicinity) on a much faster timescale than that on which the composition of the population changes. By working in an SDE (or equivalent FPE) setting that approximates the stochastic dynamics, the fast timescales can be clearly identified. On removing these fast transient dynamics, the SLVC model can be approximated by a reduced model in $M-1$ variables that is of similar form to the Moran model, and thus a precise mapping can be determined. 

Our analysis begins with a consideration of the neutral SLVC model, which we define as that in which all birth, death and competition rates are equal for all species. This setting allows us to determine a CM for the system and in turn calculate the reduced form of the neutral SLVC model. The neutral setting also serves as a useful reference case for determining the reduced form of the non-neutral SLVC model; the reduced non-neutral dynamics can be calculated using a perturbation theory around the neutral system. We find that the SLVC model maps to the Moran model with frequency-dependent selection with no conditions other than those stated in our approximation, that selection is weak and the population size is large (see Table 1). We note that for any given set of parameters in the SLVC model, a unique choice of payoff matrix in the Moran model does not exist. This is because it is the relative (rather than absolute) values of the payoff matrix that are consequential for the dynamics (see Eq.~(\ref{relative_defns})). The SLVC model maps to the Moran model with frequency-independent selection with conditions on the competition matrix that ensures that there is no frequency-dependent selection in the SLVC model. Finally, and as already stated, the SLVC model maps to the neutral Moran model with the largest number of conditions; that birth, death and competition rates are the same for each species.

Although we have used the neutral SLVC model as a reference case for our fast-variable elimination procedure, other choices of reference case are possible. In fact a similar approach to that outlined in this paper can be taken by implementing a perturbation theory around any choice of parameters that generates a CM in the LV dynamics. More generally, CMs in the LV system can also be achieved by setting $c_{ \alpha \kappa }=c_{ \beta \kappa }$ and $b_{ \alpha }-d_{ \alpha }=b_{ \beta }-d_{ \beta }$ for all $ \alpha $ and $ \beta $. Although the species are no longer identical (even in the absence of selection), a reduced form of the dynamics can still be calculated. However, breaking the symmetry between the species in this way can give rise to noise-induced selection (see, for instance, \cite{constable_2016}, where the competition terms are varied and \cite{parsons_quince_2010,lin_mig_1}, where the birth and death rates are varied in two-species systems). In this paper we have not explored these possibilities, instead focusing on developing a clean mapping between models in which neutrality is defined in the usual way. However considering the emergence of such noise-induced effects for the $M$-species SLVC model will be of great interest for further work.

In general we have stated that we expect the mapping that we have developed between the SDEs for each model to hold under the same range of validity as the approximations we have employed. From the diffusion approximation that gave rise to the SDEs we require that the number of individuals in both the SLVC model and Moran models is large (large $V$ and $N$ respectively) while from the fast-variable elimination procedure we additionally require that selection is weak (small $\epsilon$ and $s$ respectively). However, as noted in Section~\ref{sec_ins_FD}, this picture becomes more complicated in the case of frequency-dependent selection when a stable coexistence fixed point is present. In this case, as the system size increases, the diffusion approximation can become an increasingly poor predictor of the fixation properties of the system~\cite{doering_2006}. Thus although the mapping between the SDEs becomes increasingly accurate with increasing $N$ and $V$, the underlying stochastic dynamics of the systems may not converge in this limit. It is likely that this is the origin of the discrepancy between the fixation times in the SLVC model and Moran model in Figure~\ref{fig_4}. Further work is required using other approximations of the underlying master-equation, such as the WKB approximation~\cite{assaf_2010}, to determine to what extent the mapping established here remains valid with respect to fixation times. However it should be stressed that although the fixation time becomes poorly matched in this case, the mapping in terms of fixation probabilities still works very well (see Figure~\ref{fig_4}).

The mappings defined in this paper have a clear utility. Namely, the problem of calculating many stochastic quantities relating to the SLVC model (such as fixation probabilities and times) is reduced to finding a related result in the relevant Moran model together with an implementation of the mapping that we have defined in Table \ref{table_1}.  While stochasticity in the Lotka-Volterra model has received relatively little attention, there is a vast literature of results pertaining to the Moran model and its variants. In Section~\ref{sec_ins} of this paper, we have illustrated how fixation properties for the SLVC model can be derived in a straightforward way by employing the mapping to the Moran process. No doubt there are more rich and interesting behaviors that can be uncovered in a similar way. Since the SLVC model maps with the least stringent conditions to the Moran model with frequency-dependent selection, the increasing number of analytic results related to multi-strategy game theory~\cite{gokhale_2010,antal_2009} are of special note. While consideration of these numerous possible extensions is beyond the scope of this current paper, the inclusion of mutation in the SLVC model may be of particular interest. This is in part because of the large body of analytic work that exists in the population genetic~\cite{vogl_2012} and game theory literature~\cite{antal_2009} that relies on small mutation rates to gain analytic traction.

The form of the Moran model familiar to most population geneticists is that with frequency-independent selection. By construction this ignores interactions between alleles that alter each other's fitness. In terms of deriving analytic results on population genetics, understanding the frequency-independent case is clearly the first point of order. However, it is in some sense naive to imagine that nature would conform to this scenario. Frequency-dependent selection is in some sense the hallmark of an ecological system and arises continuously in biological systems. Indeed, despite conscious attempts to remove ecology from microbial experiments, frequency-dependent selection is often seen to emerge~\cite{maddamsetti_2015}. Just as these experimental findings have started to motivate theoretical studies~\cite{antal_2012}, from the perspective of theoretical ecology there is an increasing awareness that demographic stochasticity can have important consequences~\cite{gokhale_2013}.  It is our hope that the work presented here will prove to be of use to researchers in ecology, population genetics and game theory in seeing concrete parallels, distinctions and applications in each other's work.

\begin{acknowledgments}
GWAC thanks the Finnish Center for Excellence in Biological Interactions for funding. 
\end{acknowledgments}

\appendix
\section{Derivation of the Fokker-Planck equation for the Moran model}
\label{sec:Moran_setup}

In Sec.~\ref{sec:model_defn} of the main text we defined the $M$-allele Moran model with selection through the transition rates in Eqs.~(\ref{Moran_trans_sel_1})-(\ref{Moran_trans_sel_3}). We considered two types of selection. In the first type, the fitness weightings, $W_a$, were independent of the number of individuals carrying a particular allele, that is, independent of $\underline{n}$. For the second type the fitness weightings depended on the population composition in a way which was given by Eq.~(\ref{W_freq_dep}). In both cases, when these transition rates are substituted into the master equation
\begin{equation}
\frac{\mathrm{d}P_{\underline{n}}(t)}{\mathrm{d}t} = \sum_{\underline{n}' \neq \underline{n}}\left[ T(\underline{n}|\underline{n}')P(\underline{n}',t) - T(\underline{n}'|\underline{n})P(\underline{n},t)\right],
\label{master_Moran}
\end{equation}
they give the stochastic dynamics of the population. In this Appendix we will derive the FPE for the $M$-allele Moran model with selection, by applying the diffusion approximation to the master equation with these particular transition rates. 

To make the diffusion approximation we write $x_\alpha = n_\alpha/N$ and introduce the notation $F_{\alpha,\beta}(\underline{x})$ for the above transition rates for moving from state $x_\alpha$ to state $x_\alpha + N^{-1}$ and from state $x_\beta$ to state $x_\beta - N^{-1}$. In terms of the $\underline{x}$ variables, the master equation (\ref{master_Moran}) becomes, after expanding in powers of $N^{-1}$, relabelling and combining,
\begin{eqnarray}
\frac{\partial P}{\partial t} &=& - \frac{1}{N}\,\sum^{M-1}_{a \neq b} \frac{\partial }{\partial x_a} \left[ \left\{F_{a,b}\left( \underline{x} \right) - F_{b,a}\left( \underline{x} \right) \right\} P\left( \underline{x},t\right) \right] \nonumber \\
&-& \frac{1}{N}\,\sum^{M-1}_{a=1} \frac{\partial }{\partial x_a} \left[ \left\{F_{a,M}\left( \underline{x} \right) - F_{M,a}\left( \underline{x} \right) \right\} P\left( \underline{x},t\right) \right] \nonumber \\
&+& \frac{1}{2 N^2}\,\sum^{M-1}_{a \neq b} \frac{\partial^2 }{\partial x^2_a} \left[ \left\{F_{a,b}\left( \underline{x} \right) + F_{b,a}\left( \underline{x} \right) \right\} P\left( \underline{x},t\right) \right] \nonumber \\
&+& \frac{1}{2 N^2}\,\sum^{M-1}_{a=1} \frac{\partial^2 }{\partial x^2_a} \left[ \left\{F_{a,M}\left( \underline{x} \right) + F_{M,a}\left( \underline{x} \right) \right\} P\left( \underline{x},t\right) \right] \nonumber \\
&-& \frac{1}{N^2}\,\sum^{M-1}_{a \neq b} \frac{\partial^2 }{\partial x_a \partial x_b} \left[ F_{a,b}\left( \underline{x} \right) P\left( \underline{x},t\right) \right] + \mathcal{O}\left( \frac{1}{N^3} \right). \nonumber \\
\label{simplified_Moran_FPE}
\end{eqnarray}
This is the FPE for the Moran model:
\begin{eqnarray}
\frac{\partial P}{\partial t} &=& - \frac{1}{N}\,\sum^{M-1}_{a = 1} \frac{\partial }{\partial x_a} \left[ A_a \left( \underline{x}\right) P\left( \underline{x},t\right) \right] \nonumber \\
&+& \frac{1}{2 N^2}\,\sum^{M-1}_{a,b=1} \frac{\partial^2 }{\partial x_a \partial x_b} \left[ B_{a b}\left( \underline{x}\right) P\left( \underline{x},t\right) \right],
\label{Moran_FPE}
\end{eqnarray}
where
\begin{eqnarray}
A_a\left(\underline{x}\right) &=& \sum^{M-1}_{b \neq a} \left\{ \left[ F_{a,b}\left(\underline{x}\right) - F_{b,a}\left(\underline{x}\right) \right] \right.
\nonumber \\
&+& \left. \left[ F_{a,M}\left(\underline{x}\right) - F_{M,a}\left(\underline{x}\right) \right] \right\}, \nonumber \\
B_{a a}\left(\underline{x}\right) &=& \sum^{M-1}_{b \neq a} \left\{ \left[ F_{a,b}\left( \underline{x}\right) + F_{b,a}\left(\underline{x}\right) \right] \right.
\nonumber \\
&+& \left. \left[ F_{a,M}\left( \underline{x}\right) + F_{M,a}\left(\underline{x}\right) \right] \right\}, \nonumber \\
B_{a b}\left(\underline{x}\right) &=& - \left[ F_{a,b}\left( \underline{x}\right) + F_{b,a}\left(\underline{x}\right) \right] \ \ \ \left( a \neq b \right).
\label{gen_A_and_B}
\end{eqnarray}
This is as far as one can go without using the specific forms for the transition rates, so now we consider the two cases of frequency-independent selection and frequency-dependent selection in turn.

\subsection{The Fokker-Planck equation for the Moran model with frequency-independent selection}
\label{sec:Moran_setup_indep}

Before beginning the derivation, we simplify the expressions in Eqs.~(\ref{Moran_trans_sel_1})-(\ref{Moran_trans_sel_3}) by using Eq.~(\ref{selection_s}) and expanding in $s$. Since $\sum^{M}_{\gamma = 1} W_\gamma n_\gamma = N + s\,\sum^{M}_{\gamma = 1} \rho_\gamma n_\gamma$, we have 
\begin{eqnarray*}
\left[ \sum^{M}_{\gamma = 1} W_\gamma n_\gamma \right]^{-1} &=& N^{-1} \left[ 1 + 
\frac{s}{N}\,\sum^{M}_{\gamma = 1} \rho_\gamma n_\gamma \right]^{-1} \nonumber \\
&=& N^{-1} \left\{ 1 - \frac{s}{N}\,\sum^{M}_{\gamma = 1} \rho_\gamma n_\gamma 
\right. \nonumber \\
&+& \left. \frac{s^2}{N^2}\,\left(\sum^{M}_{\gamma = 1} \rho_\gamma n_\gamma\right)^2 + \mathcal{O}\left( s^3 \right) \right\}.
\end{eqnarray*}
Therefore,
\begin{eqnarray}
& & T(n_1,\ldots,n_a + 1,\ldots,n_b - 1,\ldots,n_{M-1}|\underline{n}) =
\frac{n_a}{N}\,\frac{n_b}{N} \nonumber \\
&+& s\,\frac{n_a}{N}\,\frac{n_b}{N}\,\left\{ \rho_a - 
\sum^{M}_{\gamma = 1} \rho_\gamma \frac{n_\gamma}{N} \right\} + \mathcal{O}\left( s^2 \right),
\label{indep_Moran_trans_small_s_1}
\end{eqnarray}
if $a \neq b$, 
\begin{eqnarray}
& & T(n_1,\ldots,n_a + 1,\ldots,n_{M-1}|\underline{n}) =
\frac{n_a}{N}\,\frac{N - \sum^{M-1}_{b = 1} n_b}{N} \nonumber \\
&+& s\,\frac{n_a}{N}\,\frac{N - \sum^{M-1}_{b = 1} n_b}{N}\,
\left\{ \rho_a - \sum^{M}_{\gamma = 1} \rho_\gamma \frac{n_\gamma}{N} \right\} + \mathcal{O}\left( s^2 \right),
\label{indep_Moran_trans_small_s_2}
\end{eqnarray}
and
\begin{eqnarray}
& & T(n_1,\ldots,n_a - 1,\ldots,n_{M-1}|\underline{n}) =
\frac{n_a}{N}\,\frac{N - \sum^{M-1}_{b = 1} n_b}{N} \nonumber \\
&+& s\,\frac{n_a}{N}\,\frac{N - \sum^{M-1}_{b = 1} n_b}{N}\,
\left\{ \rho_M - \sum^{M}_{\gamma = 1} \rho_\gamma \frac{n_\gamma}{N} \right\} + \mathcal{O}\left( s^2 \right).
\label{indep_Moran_trans_small_s_3}
\end{eqnarray}

In terms of the $F_{\alpha \beta}(\underline{x})$ these are, omitting terms of order $s^2$ and higher,
\begin{eqnarray}
F_{a,b}(\underline{x}) &=& x_a x_b \left\{ 1 + s\,\left[\rho_a - \sum^{M}_{\gamma = 1} \rho_\gamma x_\gamma \right] \right\}, \ \mathrm{for\ } a \neq b,
\nonumber \\
F_{a, M}(\underline{x}) &=& x_a \left( 1 - \sum^{M-1}_{b = 1} x_b \right)\,\left\{ 1 + s\,\left[\rho_a - \sum^{M}_{\gamma = 1} \rho_\gamma x_\gamma \right] \right\},
\nonumber \\
F_{M, a}(\underline{x}) &=& x_a \left( 1 - \sum^{M-1}_{b = 1} x_b \right)\,\left\{ 1 + s\,\left[\rho_M - \sum^{M}_{\gamma = 1} \rho_\gamma x_\gamma \right] \right\}.
\nonumber \\
\label{indep_Moran_Fs}
\end{eqnarray}
Using these specific forms, one finds from Eq.~(\ref{gen_A_and_B}) that
\begin{eqnarray}
A_a\left(\underline{x}\right)&=& s\sum^{M-1}_{b \neq a} x_a x_b \left( \rho_a - \rho_b \right) \nonumber \\
&+& s x_a \left( 1 - \sum^{M-1}_{b = 1} x_b \right) \left( \rho_a - \rho_M \right) + \mathcal{O}\left( s^2 \right), \nonumber \\
B_{a a}\left(\underline{x}\right)&=& 2\sum^{M-1}_{b \neq a} x_a x_b + 2 x_a \left( 1 - \sum^{M-1}_{b = 1} x_b \right) + \mathcal{O}\left( s \right), \nonumber \\
B_{a b}\left(\underline{x}\right)&=& - 2 x_a x_b + \mathcal{O}\left( s \right), \ \ \ \left( a \neq b\right).
\label{indep_explicit_Moran_A_and_B}
\end{eqnarray}

After the introduction of the new time scale $\tau=t/N$, the FPE may be written in terms of the set of $(M-1)$ SDEs (\ref{SDE_Moran}). The function $A_a(\underline{x})$ may be slightly rewritten to give Eq.~(\ref{Moran_A}). Similarly the diagonal elements of the functions $B_{a b}(\underline{x})$ may be simplified to $B_{a a}(\underline{x}) = 2x_a - 2x^2_a = 2x_a(1-x_a)$, giving Eq.~(\ref{Moran_B}).

The result for $B_{a b}(\underline{x})$ is known~\cite{kimura_1955}. One check on the form of $A_a(\underline{x})$ is to take $M=2$, so that $a$ and $b$ take only the value $1$. Then, dropping the index on $A$ and on $x$, we have that $A(x)=s \rho_1 x(1-x) - s x \rho_2 (1- x) = s(\rho_1 - \rho_2) x(1-x)$, which is the known result for two alleles, up to a constant~\cite{crow_kimura_into}. As a second, and more substantial, check we can start from the master equation with transition rates (\ref{Moran_trans_sel_1})-(\ref{Moran_trans_sel_3}) and work out the equation for $d\langle n_k \rangle/d\tau$. One finds
\begin{eqnarray}
\frac{d\langle x_k \rangle}{d\tau} &=& \sum^{M-1}_{a \neq b} \delta_{k a} \frac{W_a x_a x_b}{\sum^{m}_{\gamma = 1}W_\gamma x_\gamma} - \sum^{M-1}_{a \neq b} \delta_{k b} \frac{W_a x_a x_b}{\sum^{m}_{\gamma = 1}W_\gamma x_\gamma} \nonumber \\
&+& \sum^{M-1}_{a=1} \delta_{k a} \frac{W_a x_a}{\sum^{m}_{\gamma = 1}W_\gamma x_\gamma} \left( 1 - \sum^{M-1}_{b = 1} x_b \right) \nonumber \\
&-& \sum^{M-1}_{a=1} \delta_{k a} \frac{W_m x_a}{\sum^{m}_{\gamma = 1}W_\gamma x_\gamma} \left( 1 - \sum^{M-1}_{b = 1} x_b \right).
\label{from_master_eqn}
\end{eqnarray}
There are a number of points to be made about this equation. First, we have replaced $n_a$ by $N x_a$. Since $n_a$ always appears in the combination $n_a/N$, there was no need to take $N \to \infty$ to eliminate extra factors of $N$. Second, there should be angle brackets around all the terms on the right-hand side; these have been omitted so that the expression does not look so cluttered. This is permitted, since in the limit $N \to \infty$, the average of products is the product of the averages, that is, $\langle x_a x_b \rangle = \langle x_a \rangle \langle x_b \rangle$. In the following we will also omit the angle brackets on the left-hand side, since we are attempting to derive the macroscopic equation, which is an equation for the macroscopic variable $x_a$, written without angle brackets. Finally, we also note that the factor $\sum^M_{\beta = 1} W_\gamma x_\gamma$ is common throughout in the denominator, and so we may multiply through by it, to find
\begin{eqnarray}
& & \left( \sum^{M}_{\gamma = 1}W_\gamma x_\gamma\right)\,\frac{d x_k}{d\tau} =
W_k x_k \sum^{M-1}_{b \neq k} x_b - x_k \sum^{M-1}_{a \neq k} W_a x_a \nonumber \\
&+& W_k x_k \left( 1 - \sum^{M-1}_{b = 1} x_b \right) - W_M x_k \left( 1 - \sum^{M-1}_{b = 1} x_b \right) \nonumber \\
&=& W_k x_k \sum^{M-1}_{a = 1} x_a - x_k \sum^{M-1}_{a =1} W_a x_a \nonumber \\
&+& \left( W_k - W_M \right)\,x_k \left( 1 - \sum^{M-1}_{b = 1} x_b \right).
\label{simplification_1}
\end{eqnarray}
We now write $W_\alpha = 1 + s\rho_\alpha$, as in Eq.~(\ref{selection_s}). The terms of order $1$ on the right-hand side are seen to cancel, and so the whole of the right-hand side is of order $s$, and $W_\alpha$ may be replaced everywhere by $s\rho_\alpha$. In addition the factor in brackets on the left-hand side is $\sum^M_{\gamma = 1}x_\gamma + \mathcal{O}(s)$, and so to the order we are working at is simply equal to $1$. Therefore the macroscopic equation reads
\begin{eqnarray}
\frac{d x_k}{d\tau} &=& s \rho_k x_k \sum^{M-1}_{a = 1} x_a - s x_k \sum^{M-1}_{a=1} \rho_a x_a \nonumber \\
&+& s \left( \rho_k - \rho_M \right)\,x_k \left( 1 - \sum^{M-1}_{b = 1} x_b \right) + \mathcal{O}\left( s^2 \right) \nonumber \\
&=& s \rho_k x_k - s x_k \sum^{M-1}_{a =1} \rho_a x_a \nonumber \\
&-& s \rho_M x_k \left( 1 - \sum^{M-1}_{b = 1} x_b \right) + \mathcal{O}\left( s^2 \right), 
\label{simplification_2}
\end{eqnarray}
which has the form $d x_k/d\tau = A_k(\underline{x})$, where $A_k(\underline{x})$ is given by Eq.~(\ref{Moran_A}).

\subsection{The Fokker-Planck equation for the Moran model with frequency-dependent selection}
\label{sec:Moran_setup_dep}

In this case the average fitness of the population (see Eq.~(\ref{curly_W})) can be expressed as 
\begin{eqnarray}
\mathcal{W}(\underline{n}) &=& N \left\{  1 + \frac{s}{N^{2}} \left[ \sum_{a,b=1}^{M-1} g_{ab} n_a n_b  \right. \right. \nonumber \\ &+& \left. \left. \left(N - \sum_{b}^{M-1} n_b \right) \sum_{b=1}^{M-1} \left( g_{bM} + g_{Mb} \right) n_{b} \right. \right. \nonumber \\ &+& \left. \left. g_{MM} \left(N - \sum_{b}^{M-1} n_b \right)^{2} \right]   \right\} \,.
\label{dep_explicit_curly_W}
\end{eqnarray}
This then leads to the following results for the combinations of transition rates which are of interest to us:
\begin{eqnarray}
& & F_{ab}(\underline{x}) - F_{ba}(\underline{x}) = s x_a x_b \left[ \sum^{M-1}_{c=1} \left( g_{ac} - g_{bc} \right)x_c \right. \nonumber \\
& & \left. + \left( g_{aM} - g_{bM} \right) \left( 1 - \sum^{M-1}_{c=1} x_c \right) \right] + \mathcal{O}\left( s^2 \right) 
\label{dep_Moran_trans_small_s_1}
\end{eqnarray}
and
\begin{eqnarray}
& & F_{aM}(\underline{x}) - F_{Ma}(\underline{x}) = s x_a \left( 1 - \sum^{M-1}_{b=1} x_b \right)  \left[ \sum^{M-1}_{c=1} \left( g_{ac} - g_{Mc} \right)x_c \right. \nonumber \\
& & \left. + \left( g_{aM} - g_{MM} \right) \left( 1 - \sum^{M-1}_{c=1} x_c \right) \right] + \mathcal{O}\left( s^2 \right).
\label{dep_Moran_trans_small_s_2}
\end{eqnarray}
Substituting these expressions into Eq.~(\ref{gen_A_and_B}), one finds that 
\begin{eqnarray}
& & A_a(\underline{x}) = sx_a \left\{ \left[ g_{aM} - g_{MM} \right] + \sum^{M-1}_{b=1} \left[ g_{ab} - g_{aM} - g_{bM} \right. \right. \nonumber \\
& & \left. - g_{Mb} + 2g_{MM} \right] x_b  - \sum^{M-1}_{b,c=1} g_{bc} x_b x_c + \sum^{M-1}_{b,c=1} g_{bM} x_b x_c \nonumber \\
& & \left. + \sum^{M-1}_{b,c=1} g_{Mb} x_b x_c - \sum^{M-1}_{b,c=1} g_{MM} x_b x_c \right\},
\label{dep_Moran_trans_small_s_3}
\end{eqnarray}
with the form of $B_ (\underline{x})$ being the same as in the frequency-independent case, given in Eq.~(\ref{indep_explicit_Moran_A_and_B}). Now interchanging the $b$ and $c$ labels in the sum involving $g_{Mb}x_bx_c$ in Eq.~(\ref{dep_Moran_trans_small_s_3}), and introducing $\mathcal{G}_{aM}=g_{aM}-g_{MM}$ and $G_{ab} = g_{ab} - g_{aM} - g_{Mb} + g_{MM}$ one obtains Eq.~(\ref{Replicator_A}) in the main text.

As for the case of frequency-independent selection, we can check this result for $A_a(\underline{x})$ by calculating $d\langle n_k \rangle/d\tau$ directly from the master equation. The steps leading to Eq.~(\ref{simplification_1}) still hold --- as long as $W_{\alpha}$ is replaced by $\langle W_{\alpha}(\underline{x}) \rangle$, but as in the frequency-independent case, since we are deriving what is the macroscopic equation this is simply $W(\langle \underline{x} \rangle )$, that is, $W(\underline{x})$. The comments below Eq.~(\ref{simplification_1}) also hold here, and substitution of the order $s$ term in Eq.~(\ref{W_freq_dep}) does indeed give $d x_k/d\tau = A_k(\underline{x})$, where $A_k(\underline{x})$ is given by Eq.~(\ref{dep_Moran_trans_small_s_3}).

\section{Using results from the Moran model to calculate fixation quantities in the SLVC}\label{app_fixprob}

In this appendix we will use results on fixation probabilities and times in the Moran model to calculate equivalent quantities in the SLVC model. In the first section, we will consider the neutral model. Here we will calculate the probability of fixation of any species and the probability of any series of extinctions, as well as the unconditional mean time until each successive extinction. In the following two sections we will derive results for the fixation of alleles in the frequency-independent model and frequency-dependent model respectively. In particular, in these later two cases we shall calculate the fixation probability of alleles in degenerate scenarios in which there are $M$ species but only two selection strengths (frequency-independent selection) or two competition regimes (frequency-dependent selection).

\subsection{The neutral models}

We first calculate the probability of a particular sequence of extinctions. Consider an $M$-allele single locus neutral haploid Moran model with the frequency of each allele denoted $z_a$, $a=1,\ldots M-1$. We wish to calculate the fixation probability of a particular allele, say allele $b$, with frequency $z_b$. We begin by noting that if we are only interested in the dynamics of a single allele, we can group together the remaining alleles and treat these as a single type. The frequency of the remaining type, which we shall denote $z'_r$ is then simply given by $z'_r = 1 - z_b$. In a two allele system, the fixation probability of an allele is simply equal to its relative initial frequency in the population~\cite{crow_kimura_into};
\begin{eqnarray*}
P_{\mathrm{Fix}}^{\{b\} }  = \frac{ z^{(0)}_{b} }{ z^{(0)}_{b} + z'^{(0)}_r} = z^{(0)}_{b} \,.
\end{eqnarray*}
This is the probability that allele $b$ fixates, or equivalently, that $b$ does not go to extinction. 

We next consider the dynamics within the subpopulation of frequency $z'_r$. We ask what is the fixation probability of an allele $c$ (of frequency $z^{(0)}_{c}$ in the global population) within the subpopulation. This is given by
\begin{eqnarray*}
P_{\mathrm{Fix}}^{\{c\}|(\mathrm{subpop.\,}1)} = \frac{ z^{(0)}_c }{ z'^{(0)}_r } = \frac{ z^{(0)}_{c} }{ 1 - z^{(0)}_{b} } \,.
\end{eqnarray*}
The probability that $c$ fixates first within the subpopulation, followed by $b$ fixating in the global population is then 
\begin{eqnarray*}
P_{\mathrm{Fix}}^{b,c} &=& P_{ \mathrm{Fix} }^{ \{b\} } P_{ \mathrm{Fix} }^{ \{c\}|(\mathrm{subpop.\,}1) } \, \\
				 &=& z^{(0)}_{b} \frac{ z^{(0)}_{c} }{ 1 - z^{(0)}_{b} } \,.
\end{eqnarray*}
Equivalently, this is the probability that allele $c$ is the final allele to become extinct, while allele $b$ does not go to extinction.  

Iterating this argument over successive subpopulations, the probability of a given sequence $\underline{S}=(\alpha_M, \alpha_{M-1},\ldots,\alpha_2)$ of extinctions is~\cite{baxter_2007}
\begin{eqnarray*}
P_{\mathrm{Fix}}^{\underline{S}} = \prod^{M-1}_{a=1} \frac{ z^{(0)}_{\alpha_a} }{ 1 - \sum_{b=1}^{a-1} z^{(0)}_{\alpha_b}  } \,, \label{eq_fixprob_neutral_z} 
\end{eqnarray*}
where $\alpha_M$ is the first to go extinct, $\alpha_{M-1}$ the second,..., until only allele $\alpha_1$ remains. Our final step is to simply transform this into a function of the original $\bm{x}^{(0)}$ initial condition variables using Eq.~(\ref{IC}). This leads to Eq.~(\ref{eq_fixprob_SLVC_neutral}) of the main text.

We next calculate the unconditional mean time to fixation of the $r^{\rm th}$ allele. In the Moran model with the frequency of each allele denoted by $z_a$, this is given by $\langle T_{(r)}^{\mathrm{(M)}}(\bm{z}^{(0)}) \rangle$~\cite{baxter_2007}
\begin{eqnarray}
\langle T_{(r)}^{\mathrm{(M)}}(\bm{z}^{(0)}) \rangle &=& - N^{2} \sum_{a=r}^{M-1} (-1)^{a-r} \binom{a-1}{r-1} \times \nonumber \\
      & & \sum_{ \bm{\alpha}} \left(\sum_{b=1}^{a} z^{(0)}_{\alpha_{b}} \right)\ln\left[\sum_{b=1}^{a} z^{(0)}_{\alpha_{b}} \right] \,. \label{eq_fixtime_neutral_moran}
\end{eqnarray}
Here the summation over $\bm{\alpha}=\{ \alpha_1,\alpha_2,\ldots,\alpha_a\}$ denotes summation over all possible subsets of the set of positive integers $\{ 1,2,\ldots,M\}$ with $a$ entries. For instance, for $M=3$ and $a=2$
\begin{eqnarray}
&\sum_{ \bm{\alpha}} \left( \sum_{b=1}^{2} z^{(0)}_{\alpha_{b}} \right) \ln\left[\sum_{b=1}^{2} z^{(0)}_{\alpha_{b}} \right] = \nonumber \\ &(z^{(0)}_{1} + z^{(0)}_{2}) \ln\left[ z^{(0)}_{1} + z^{(0)}_{2} \right] + (z^{(0)}_{1} + z^{(0)}_{3}) \ln\left[ z^{(0)}_{1} + z^{(0)}_{3} \right]  \nonumber \\ &+ (z^{(0)}_{2} + z^{(0)}_{3}) \ln\left[ z^{(0)}_{2} + z^{(0)}_{3} \right] \,. 
\end{eqnarray}
We also note that the result Eq.~(\ref{eq_fixtime_neutral_moran}) differs slightly from that given in \cite{constable_2015} in that we have stated it in natural time units of the Moran model (that is in $t^{(\mathrm{M})}$, see Table~\ref{table_1}), rather than rescaled time, which accounts for the $N^{2}$ prefactor in our description.

Finally we require that Eq.~(\ref{eq_fixtime_neutral_moran}) is given in units and variables appropriate for the untransformed SLVC model. Using Table~\ref{table_1}, we can transform each of the initial conditions in Eq.~(\ref{eq_fixtime_neutral_moran}) into their equivalent values in the SLVC model formulation (see also Eq.~(\ref{IC})) as well as rescaling time into the natural units of the Moran model $t^{(\mathrm{LV})}$. Recalling that the average population size in the SLVC model at carrying capacity is denoted $\langle N^{(\mathrm{LV})}\rangle$, we find that the time to fixation in the SLVC model in its natural units is given by Eq.~(\ref{eq_fixtime_SLVC_neutral}).

\subsection{The models with frequency-independent selection}

We begin by recalling the dynamics of the two-allele Moran model, in which allele $1$ has fitness $W_1=1+s\rho_1$ and allele $2$ has fitness $W_2=1+s \rho_2$, that is, the $M=2$ version of the model described in Sec.~\ref{sec_moran_FI_model}. The drift and diffusion terms for this system are respectively
\begin{eqnarray}
A(z_1) &=& s (\rho_{1} - \rho_{2})z_1( 1 - z_1 ) \,, \nonumber \\ 
B(z_{1}) &=& 2 z_1(1-z_1) \,,
\end{eqnarray}
to the order we are working in $s$. The fixation probability of allele $1$ can be obtained by solving the backward FPE~\cite{gardiner_2009,constable_thesis}, and is given by
\begin{eqnarray}
P_{\mathrm{Fix}}^{\{1\} } = \frac{ 1 - \exp \left[ -sN (\rho_1 - \rho_2 )  z_1 \right] }{ 1 - \exp \left[ -sN (\rho_1 - \rho_2 ) \right] } \,.\label{eq_fi_moran_fixprob_1}
\end{eqnarray}
The fixation probability of allele $2$ is then $P_{\mathrm{Fix}}^{\{2\} }  = 1 - P_{\mathrm{Fix}}^{\{1\} } $.

We now move on to the $M$-allele Moran model, but where the $M$ alleles are acted on by only two distinct selection pressures, $\bar{\rho}_1$ and $\bar{\rho}_2$. Suppose that $\theta$ of the alleles are acted upon by selection pressure $\bar{\rho}_1$. We choose to label these $1,\ldots,\theta$. Then those labelled $\theta +1, \ldots,M$ are acted under selection pressure $\bar{\rho}_2$, that is,
\begin{eqnarray}
\rho_\alpha &=& \bar{\rho}_1 \quad \alpha =1,\dots, \theta \,, \nonumber \\
 \rho_\alpha &=& \bar{\rho}_2 \quad \alpha = \theta +1, \ldots, M\,.
\end{eqnarray}
Therefore the first $\theta$ alleles can be said to constitute subpopulation $1$, while final $M-\theta$ alleles can be designated subpopulation $2$. Since the dynamics within each subpopulation are neutral, we can say that the probability of each subpopulation fixating is simply given by $P_{\mathrm{Fix}}^{\{1\} }$ and $P_{\mathrm{Fix}}^{\{2\} }$ (see Eq~(\ref{eq_fi_moran_fixprob_1})). The reason that this can be done, is that different alleles in each subpopulation only differ by the labels given to them; we can therefore choose to label them as only belonging to a particular subpopulation, without changing the dynamics. Meanwhile, again since the subpopulations are neutral, the probability that each allele fixates within its respective subpopulation is simply equal to its initial frequency within the subpopulation. Therefore, the probability that any allele $\alpha$ fixates is equal to the product of the probability that it fixates within its subpopulation and the probability that its subpopulation fixates;
\begin{eqnarray}
P_{\mathrm{Fix}}^{\{\alpha\} } &=& \frac{1 - \exp \left[ -sN (\bar{\rho_1} - \bar{\rho_2} ) \sum_{b=1}^{\theta} z^{(0)}_{b} \right]}{1 -\exp \left[ -sN (\bar{\rho_1} - \bar{\rho_2} ) \right]} \times \nonumber \\
& & \frac{ z^{(0)}_{\alpha} }{ \sum_{b=1}^{\theta} z^{(0)}_{b}  } \, ,  \qquad \qquad \qquad \qquad \qquad 1 \leq \alpha \leq \theta \,,\nonumber \\
P_{\mathrm{Fix}}^{\{\alpha\} } &=& \left\{ 1 - \frac{1 - \exp \left[ -sN (\bar{\rho_1} - \bar{\rho_2} ) \sum_{b=1}^{\theta} z^{(0)}_{b} \right]}{1 -\exp \left[ -sN (\bar{\rho_1} - \bar{\rho_2} ) \right]} \right\} \times \nonumber \\ 
& & \frac{ z^{(0)}_{\alpha} }{\left[ 1 - \sum_{b=1}^{\theta} z^{(0)}_{b} \right]} \, , \qquad \quad  \qquad \qquad \theta < \alpha \leq M\,,
\nonumber \\
\end{eqnarray}
where the initial frequency of the $M^{\rm th}$ allele, $z^{(0)}_M$ is understood to be given by $1 - \sum^{M-1}_{b=1} z^{(0)}_b$. Notice that unlike in the neutral case, we cannot determine the probability of allele extinctions in a particular order. This is because we can only break this non-neutral system down into two distinct subpopulations, whereas in the neutral case we could break the system down into any focal allele, plus the remainder of the population.

The final task is to write the above equations in terms of the original SLVC model variables and parameters. Note that the mapping for the parameter $\rho_a$ depends on a combination of the parameters in terms of the original SLVC model variables (see Table~\ref{table_1}). In order to make use of the calculation above, we require that each $\rho_\alpha$ for $\alpha\leq\theta$ is identical, however this does not mean that the parameters in the SLVC model need be identical; we merely require that 
$ b_0\beta_\alpha - d_0 \delta_\alpha - (b_0 - d_0) \gamma_{\alpha M} = \bar{\rho}_1 $ for $\alpha\leq\theta$  and $ b_0\beta_\alpha - d_0 \delta_\alpha - (b_0 - d_0) \gamma_{\alpha M} = \bar{\rho}_2 $ for $\alpha>\theta$. This leads to the equivalent expressions given by Eq.~(\ref{parameter_map_1}) for the model for the model under consideration, which can be seen clearly in the parameters used to generate figure~\ref{fig_4} (see Appendix~\ref{app_params}). The fixation probabilities in SLVC model notation are then given by Eq.~(\ref{eq_fixprob_SLVC_indep}) of the main text.

\subsection{The models with frequency-dependent selection}

Let us begin by recalling the dynamics of the two-strategy model, that is, the $M=2$ version of the model described in Sec.~\ref{sec_moran_FD_model}:
\begin{eqnarray}
A(z_1) &=& s z_1( 1 - z_1 )(\mathcal{G}_{12} + G_{11} z_1) \,, \nonumber \\
 B(z_{1}) &=& 2 z_1(1-z_1) \,,
\end{eqnarray}
to the order we are working in $s$. The fixation probability of allele $1$ can be obtained by solving the backward FPE~\cite{gardiner_2009,constable_thesis}. Defining the function
\begin{equation}
l(z^{(0)}_{1}) = \sqrt{ \frac{ s N }{2 |G_{11}| } } \left( \mathcal{G}_{12}
+ G_{11} z^{(0)}_{1} \right) \,,
\end{equation}
one finds
\begin{equation}
P_{\mathrm{Fix}}^{\{1\} } = \frac{1 - \chi[l(z^{(0)}_1)]}{1 - \chi[l(1)]}, 
\label{eq_fd_moran_fixprob_1}
\end{equation}
where
\begin{eqnarray}
\chi[l(z^{(0)}_1)] = \frac{\mathrm{erfi}\left[ l(z^{(0)}_{1}) \right] }{\mathrm{erfi}\left[ l(0) \right] } \,, \quad \rm{if} \quad G_{11} > 0 \,,\nonumber \\
\chi[l(z^{(0)}_1)] = \frac{\mathrm{erfc}\left[ l(z^{(0)}_{1}) \right]  }{\mathrm{erfc}\left[ l(0) \right] } \,, \quad \rm{if} \quad G_{11} < 0 \,. \label{eq_chi_1}
\end{eqnarray}
Here erfc and erfi are the complementary and imaginary error functions respectively~\cite{handbook_1965,HTFS_1953}. Again, the fixation probability of strategy $2$ is $P_{\mathrm{Fix}}^{\{2\} }  = 1 - P_{\mathrm{Fix}}^{\{1\} } $. 

In a similar fashion to the frequency-independent case, we now envisage a scenario in which we have multiple strategies, but the payoff matrix can be partitioned such that;
\begin{eqnarray}
g_{\alpha \beta} &= \bar{g}_{11} \,, \quad \alpha,\beta \leq \theta \,; \quad g_{\alpha \beta} = \bar{g}_{12} \,, \quad \alpha \leq \theta < \beta \,; \nonumber \\
g_{\alpha \beta} &= \bar{g}_{22} \,, \quad \alpha, \beta > \theta \,; \quad g_{\alpha \beta} = \bar{g}_{21} \,, \quad \beta \leq \theta < \alpha \,. \nonumber 
\end{eqnarray}

Let the first $\theta$ alleles be subpopulation $1$, while final $M-\theta$ alleles can be designated population two. Since the dynamics within each subpopulation are neutral, we can say that the probability of each subpopulation fixating is simply given by $P_{\mathrm{Fix}}^{\{1\} }$ (see Eq~(\ref{eq_fd_moran_fixprob_1})) and $P_{\mathrm{Fix}}^{\{2\} } = 1 - P_{\mathrm{Fix}}^{\{1\} }$. Meanwhile the probability that each allele fixates within its respective subpopulation is simply equal to its initial frequency within the subpopulation. Therefore, the probability that any allele $\alpha$ fixates is equal to the product of the probability that it fixates within its subpopulation and the probability that its subpopulation fixates:
\begin{eqnarray}
P_{\mathrm{Fix}}^{\{\alpha\} } &=& \frac{ 1 - \chi\left[ l(\bm{z}^{(0)}) \right]  }{ 1 - \chi\left[ l(1) \right] } \frac{ z^{(0)}_{\alpha} }{ \sum_{b=1}^{\theta} z^{(0)}_{b}  } \,, \quad \rm{if} \quad 1 \leq \alpha \leq \theta\,,\nonumber \\
P_{\mathrm{Fix}}^{\{\alpha\} } &=& \frac{ 1 - \chi\left[ l(\bm{z}^{(0)}) \right]  }{ 1 - \chi\left[ l(1) \right] } \frac{ z^{(0)}_{\alpha} }{ 1 - \sum_{b=1}^{\theta} z^{(0)}_{b}} \,, \quad \rm{if} \quad \theta < \alpha \leq M \,,\nonumber \\
\end{eqnarray}
where
\begin{eqnarray}
\chi[l(\bm{z}^{(0)})] = \frac{\mathrm{erfi}\left[ l(\bm{z}^{(0)}) \right] }{\mathrm{erfi}\left[ l(0) \right] } \,, \quad \rm{if} \quad \bar{G}_{11} > 0 \,,\nonumber \\
\chi[l(\bm{z}^{(0)})] = \frac{\mathrm{erfc}\left[ l(\bm{z}^{(0)}) \right]  }{\mathrm{erfc}\left[ l(0) \right] } \,, \quad \rm{if} \quad \bar{G}_{11} < 0 \,, \label{eq_chi_M}
\end{eqnarray}
and where
\begin{equation}
l(\bm{z}^{(0)}) = \sqrt{ \frac{ s N }{2 |\bar{G}_{11}| } } \left( \bar{\mathcal{G}}_{12} + \bar{G}_{11} \sum_{b=1}^{\theta} z^{(0)}_{b} \right) \,.
\end{equation}
Here, by $l(1)$ and $l(0)$ we mean the value of $l$ when $\sum_{b=1}^{\theta} z^{(0)}_{b} =1$ and $\sum_{b=1}^{\theta} z^{(0)}_{b} = 0$, respectively. Finally, writing this as a function of the SLVC variables and parameters, we obtain Eq.~(\ref{eq_fixprob_SLVC_dep}) of the main text.

\section{Parameters used in figures}\label{app_params}

In Figure \ref{fig_1}, the parameters used for the SLVC are 
\begin{eqnarray}
M&=3 \,, \quad b_{0}& = 3 \,, \,\, \quad d_{0} = 2 \, , \nonumber \\
c_{0}&=0.1 \,, \quad V&=100 \,, \quad \epsilon =0 \,.
\end{eqnarray}

In Figure \ref{fig_2}, the parameters used for the SLVC are 
\begin{eqnarray}
M&=5 \,, \quad b_{0}& = 3 \,, \,\, \quad d_{0} = 1 \, , \nonumber \\
c_{0}&=0.1 \,, \quad V&=20 \,, \quad \epsilon =0 \,,
\end{eqnarray}
and the parameters used for the Moran model are
\begin{equation}
N=200 \,.
\end{equation}
The initial conditions used in the SLVC model and Moran model are respectively 
\begin{align}
\bm{x}^{\mathrm{(0,LV)}} = 10\left( \begin{array}{c} \frac{2}{3} \kappa \\ \frac{1}{3} \kappa \\ \frac{3}{6}(1 - \kappa )  \\ \frac{2}{6}(1 - \kappa )  \\ \frac{1}{6}(1 - \kappa ) \end{array} \right), \ \ 
\bm{x}_{0}^{\mathrm{(0,M)}} = \left( \begin{array}{c} \frac{2}{3} \kappa \\ \frac{1}{3} \kappa \\ \frac{3}{6}(1 - \kappa )  \\ \frac{2}{6}(1 - \kappa )  \\ \frac{1}{6}(1 - \kappa ) \end{array} \right) \,.
\end{align}

In Figure \ref{fig_3}, the parameters used for the SLVC are 
\begin{eqnarray}
M&=6 \,, \quad b_{0}& = 2 \,, \,\, \quad d_{0} = 1 \, , \nonumber \\
c_{0}&=0.1 \,, \quad V&=10 \,, \quad \epsilon =0.04 \,, 
\end{eqnarray}
\begin{align}
\bm{\beta} = \left( \begin{array}{c} 1 \\ 2 \\ 0.5 \\ -1  \\ 1 \\ 0 \end{array} \right), \ \ 
\bm{\delta} = \left( \begin{array}{c} 0 \\ 2 \\ -1  \\ -2   \\ 2 \\ 0 \end{array} \right), \ \ 
\end{align}
\begin{align}
 \begin{array}{c} \gamma_{16} = 1 \,, \\ \gamma_{26} = 1 \,, \\ \gamma_{36} = 1 \,, \\ \gamma_{46} = 0 \,,  \\ \gamma_{56} = 0 \,, \\ \gamma_{66} = 0 \,,  \end{array} \ \  
\begin{array}{c} \gamma_{61} = -1 \,, \\ \gamma_{62} = 0.4 \,, \\ \gamma_{63} = 0.2 \,, \\ \gamma_{64} = 0 \,,  \\ \gamma_{65} = 0 \,, \\ \vphantom{0}  \end{array},\ \ 
\end{align}
and the parameters used for the Moran model are
\begin{equation}
N=100 \,, \qquad s = 0.02 \,,
\end{equation}
\begin{align}
\bm{\rho} = \left( \begin{array}{c} 1 \\ 1 \\ 1 \\ 0  \\ 0 \\ 0 \end{array} \right) \,. 
\end{align}
Note that these choices of parameters make the models mappable to each other (see Table \ref{table_1}). Also note that although the values of the parameters $\gamma_{Ma}$ are needed to specify the dynamics of the SLVC simulation, they are not required for the mapping, and thus the same mapping between the SLVC model and the Moran model with frequency-independent selection holds for any order one choice for the parameters $\gamma_{Ma}$. The initial conditions used in the SLVC model and Moran model are respectively 
\begin{align}
\bm{x}_{0}^{\mathrm{(0,LV)}} = 10\left( \begin{array}{c} \frac{1}{6} \kappa \\ \frac{2}{6} \kappa \\ \frac{3}{6} \kappa \\ \frac{1}{6}(1 - \kappa )  \\ \frac{2}{6}(1 - \kappa )  \\ \frac{3}{6}(1 - \kappa ) \end{array} \right) \, , \ \ 
\bm{x}_{0}^{\mathrm{(0,M)}} = \left( \begin{array}{c} \frac{1}{6} \kappa \\ \frac{2}{6} \kappa \\ \frac{3}{6} \kappa \\ \frac{1}{6}(1 - \kappa )  \\ \frac{2}{6}(1 - \kappa )  \\ \frac{3}{6}(1 - \kappa ) \end{array} \right) \,.
\end{align}

In Figure \ref{fig_4}, the parameters used for the SLVC are
\begin{eqnarray}
M&=4 \,, \quad b_{0}& = 2 \,, \,\, \quad d_{0} = 1 \, , \nonumber \\
c_{0}&=0.1 \,, \quad V&=20 \,, \quad \epsilon =0.03 \,,
\end{eqnarray}
\begin{align}
\bm{\beta} = \left( \begin{array}{c} 1 \\ 1 \\ 0 \\ 0 \end{array} \right), \ \ 
\bm{\delta} = \left( \begin{array}{c} 0 \\ 0 \\ 1  \\ 1 \end{array} \right), \ \ 
\end{align}
\begin{align}
\gamma = \left( \begin{array}{cccc} 4 & 4 & -1 & -1 \\  4 & 4 & -1 & -1 \\ -4 & -4 & 1 & 1 \\ -4 & -4 & 1 & 1 \end{array} \right) \,,
\end{align}
and the parameters used for the Moran model are
\begin{equation}
N=200 \,, \qquad s = 0.015 \,,
\end{equation}
\begin{align}
g = \left( \begin{array}{cccc} -6 & -6 & 3 & 3 \\  -6 & -6 & 3 & 3 \\ -1 & -1 & -2 & -2 \\ -1 & -1 & -2 & -2 \end{array} \right) \,.
\end{align}
Note that these choices of parameters make the models mappable to each other (see Table \ref{table_1}). The initial conditions used in the SLVC model and Moran model are respectively 
\begin{align}
\bm{x}_{0}^{\mathrm{(0,LV)}} = 10\left( \begin{array}{c} \frac{2}{3} \kappa \\ \frac{1}{3} \kappa \\  \frac{1}{3}(1 - \kappa )  \\ \frac{2}{3}(1 - \kappa )  \end{array} \right) \, , \ \ 
\bm{x}_{0}^{\mathrm{(0,M)}} =  \left( \begin{array}{c} \frac{2}{3} \kappa \\ \frac{1}{3} \kappa \\  \frac{1}{3}(1 - \kappa )  \\ \frac{2}{3}(1 - \kappa )  \end{array} \right) \, .
\end{align}

In Figure \ref{fig_5}, the parameters used for the SLVC are 
\begin{eqnarray}
M&=4 \,, \quad b_{0}& = 2 \,, \,\, \quad d_{0} = 1 \, , \nonumber \\
c_{0}&=0.1 \,, \quad V&=10 \,, \quad \epsilon =0.02 \,,
\end{eqnarray}
\begin{align}
\bm{\beta} = \left( \begin{array}{c} 1 \\ -1 \\ 0 \\ 3 \end{array} \right), \ \ 
\bm{\delta} = \left( \begin{array}{c} 0 \\ -2 \\ 0  \\ 1 \end{array} \right), \ \ 
\end{align}
\begin{align}
\gamma = \left( \begin{array}{cccc} 1 & 0 & 0 & -1 \\  0 & -0.5 & -1 & 1 \\ 0.5 & 1 & 0 & 1 \\ 1 & 0.5 & -1 & 0 \end{array} \right) \,,
\end{align}
and the parameters used for the Moran model are
\begin{equation}
N=100 \,, \qquad s = 0.01 \,,
\end{equation}
\begin{align}
g = \left( \begin{array}{cccc} 1 & 0.75 & 0 & 3 \\  0 & -0.75 & -1 & -1 \\ -0.5 & -2.25 & -2 & -1 \\ 4 & 3.25 & 4 & 5 \end{array} \right) \,.
\end{align}
Note that these choices of parameters make the models mappable to each other (see Table \ref{table_1}). The initial conditions used in the SLVC model and Moran model are respectively 
\begin{align}
\bm{x}_{0}^{\mathrm{(0,LV)}} = 10\left( \begin{array}{c} \frac{2}{3} \kappa \\ \frac{1}{3} \kappa \\  \frac{1}{3}(1 - \kappa )  \\ \frac{2}{3}(1 - \kappa )  \end{array} \right) \, , \ \ 
\bm{x}_{0}^{\mathrm{(0,M)}} =  \left( \begin{array}{c} \frac{2}{3} \kappa \\ \frac{1}{3} \kappa \\  \frac{1}{3}(1 - \kappa )  \\ \frac{2}{3}(1 - \kappa )  \end{array} \right) \, .
\end{align}

\vspace{2cm}


\begin{thebibliography}{45}%
\makeatletter
\providecommand \@ifxundefined [1]{%
 \@ifx{#1\undefined}
}%
\providecommand \@ifnum [1]{%
 \ifnum #1\expandafter \@firstoftwo
 \else \expandafter \@secondoftwo
 \fi
}%
\providecommand \@ifx [1]{%
 \ifx #1\expandafter \@firstoftwo
 \else \expandafter \@secondoftwo
 \fi
}%
\providecommand \natexlab [1]{#1}%
\providecommand \enquote  [1]{``#1''}%
\providecommand \bibnamefont  [1]{#1}%
\providecommand \bibfnamefont [1]{#1}%
\providecommand \citenamefont [1]{#1}%
\providecommand \href@noop [0]{\@secondoftwo}%
\providecommand \href [0]{\begingroup \@sanitize@url \@href}%
\providecommand \@href[1]{\@@startlink{#1}\@@href}%
\providecommand \@@href[1]{\endgroup#1\@@endlink}%
\providecommand \@sanitize@url [0]{\catcode `\\12\catcode `\$12\catcode
  `\&12\catcode `\#12\catcode `\^12\catcode `\_12\catcode `\%12\relax}%
\providecommand \@@startlink[1]{}%
\providecommand \@@endlink[0]{}%
\providecommand \url  [0]{\begingroup\@sanitize@url \@url }%
\providecommand \@url [1]{\endgroup\@href {#1}{\urlprefix }}%
\providecommand \urlprefix  [0]{URL }%
\providecommand \Eprint [0]{\href }%
\providecommand \doibase [0]{http://dx.doi.org/}%
\providecommand \selectlanguage [0]{\@gobble}%
\providecommand \bibinfo  [0]{\@secondoftwo}%
\providecommand \bibfield  [0]{\@secondoftwo}%
\providecommand \translation [1]{[#1]}%
\providecommand \BibitemOpen [0]{}%
\providecommand \bibitemStop [0]{}%
\providecommand \bibitemNoStop [0]{.\EOS\space}%
\providecommand \EOS [0]{\spacefactor3000\relax}%
\providecommand \BibitemShut  [1]{\csname bibitem#1\endcsname}%
\let\auto@bib@innerbib\@empty
\bibitem [{\citenamefont {Roughgarden}(1979)}]{roughgarden_1979}%
  \BibitemOpen
  \bibfield  {author} {\bibinfo {author} {\bibfnamefont {J.}~\bibnamefont
  {Roughgarden}},\ }\href@noop {} {\emph {\bibinfo {title} {{T}heory of
  {P}opulation {G}enetics and {E}volutionary {E}cology: {A}n {I}ntroduction}}}\
  (\bibinfo  {publisher} {Macmillan},\ \bibinfo {address} {New York},\ \bibinfo
  {year} {1979})\BibitemShut {NoStop}%
\bibitem [{\citenamefont {Moran}(1957)}]{moran_1957}%
  \BibitemOpen
  \bibfield  {author} {\bibinfo {author} {\bibfnamefont {P.~A.~P.}\
  \bibnamefont {Moran}},\ }\href@noop {} {\bibfield  {journal} {\bibinfo
  {journal} {Math. Proc. Cam. Phil. Soc.}\ }\textbf {\bibinfo {volume} {54}},\
  \bibinfo {pages} {60} (\bibinfo {year} {1957})}\BibitemShut {NoStop}%
\bibitem [{\citenamefont {Fisher}(1930)}]{fisher_1930}%
  \BibitemOpen
  \bibfield  {author} {\bibinfo {author} {\bibfnamefont {R.~A.}\ \bibnamefont
  {Fisher}},\ }\href@noop {} {\emph {\bibinfo {title} {The Genetical Theory of
  Natural Selection}}}\ (\bibinfo  {publisher} {Clarendon Press},\ \bibinfo
  {address} {Oxford},\ \bibinfo {year} {1930})\BibitemShut {NoStop}%
\bibitem [{\citenamefont {Hofbauer}\ and\ \citenamefont
  {Sigmund}(1998)}]{hofbauer_1998}%
  \BibitemOpen
  \bibfield  {author} {\bibinfo {author} {\bibfnamefont {J.}~\bibnamefont
  {Hofbauer}}\ and\ \bibinfo {author} {\bibfnamefont {K.}~\bibnamefont
  {Sigmund}},\ }\href@noop {} {\emph {\bibinfo {title} {{E}volutionary {G}ames
  and {P}opulation {D}ynamics}}}\ (\bibinfo  {publisher} {Cambridge University
  Press},\ \bibinfo {address} {Cambridge},\ \bibinfo {year} {1998})\BibitemShut
  {NoStop}%
\bibitem [{\citenamefont {Pielou}(1977)}]{pielou_1977}%
  \BibitemOpen
  \bibfield  {author} {\bibinfo {author} {\bibfnamefont {E.~C.}\ \bibnamefont
  {Pielou}},\ }\href@noop {} {\emph {\bibinfo {title} {{M}athematical
  {E}cology}}}\ (\bibinfo  {publisher} {Wiley},\ \bibinfo {address} {New
  York},\ \bibinfo {year} {1977})\BibitemShut {NoStop}%
\bibitem [{\citenamefont {Halliburton}(2004)}]{halliburton_2004}%
  \BibitemOpen
  \bibfield  {author} {\bibinfo {author} {\bibfnamefont {R.}~\bibnamefont
  {Halliburton}},\ }\href@noop {} {\emph {\bibinfo {title} {{I}ntroduction to
  {P}opulation {G}enetics}}}\ (\bibinfo  {publisher} {Pearson Press},\ \bibinfo
  {address} {New Jersey},\ \bibinfo {year} {2004})\BibitemShut {NoStop}%
\bibitem [{\citenamefont {Nowak}\ \emph {et~al.}(2004)\citenamefont {Nowak},
  \citenamefont {Sasaki}, \citenamefont {Taylor},\ and\ \citenamefont
  {Fudenberg}}]{nowak_2004}%
  \BibitemOpen
  \bibfield  {author} {\bibinfo {author} {\bibfnamefont {M.~A.}\ \bibnamefont
  {Nowak}}, \bibinfo {author} {\bibfnamefont {A.}~\bibnamefont {Sasaki}},
  \bibinfo {author} {\bibfnamefont {C.}~\bibnamefont {Taylor}}, \ and\ \bibinfo
  {author} {\bibfnamefont {D.}~\bibnamefont {Fudenberg}},\ }\href@noop {}
  {\bibfield  {journal} {\bibinfo  {journal} {Nature}\ }\textbf {\bibinfo
  {volume} {428}},\ \bibinfo {pages} {646} (\bibinfo {year}
  {2004})}\BibitemShut {NoStop}%
\bibitem [{\citenamefont {Traulsen}\ \emph {et~al.}(2005)\citenamefont
  {Traulsen}, \citenamefont {Claussen},\ and\ \citenamefont
  {Hauert}}]{traulsen_2005}%
  \BibitemOpen
  \bibfield  {author} {\bibinfo {author} {\bibfnamefont {A.}~\bibnamefont
  {Traulsen}}, \bibinfo {author} {\bibfnamefont {J.~C.}\ \bibnamefont
  {Claussen}}, \ and\ \bibinfo {author} {\bibfnamefont {C.}~\bibnamefont
  {Hauert}},\ }\href@noop {} {\bibfield  {journal} {\bibinfo  {journal} {Phys.
  Rev. Lett.}\ }\textbf {\bibinfo {volume} {95}},\ \bibinfo {pages} {238701}
  (\bibinfo {year} {2005})}\BibitemShut {NoStop}%
\bibitem [{\citenamefont {Traulsen}\ \emph {et~al.}(2008)\citenamefont
  {Traulsen}, \citenamefont {Shoresh},\ and\ \citenamefont
  {Nowak}}]{traulsen_2008}%
  \BibitemOpen
  \bibfield  {author} {\bibinfo {author} {\bibfnamefont {A.}~\bibnamefont
  {Traulsen}}, \bibinfo {author} {\bibfnamefont {N.}~\bibnamefont {Shoresh}}, \
  and\ \bibinfo {author} {\bibfnamefont {M.~A.}\ \bibnamefont {Nowak}},\
  }\href@noop {} {\bibfield  {journal} {\bibinfo  {journal} {Bull. Math.
  Biol.}\ }\textbf {\bibinfo {volume} {70}},\ \bibinfo {pages} {1410} (\bibinfo
  {year} {2008})}\BibitemShut {NoStop}%
\bibitem [{\citenamefont {Kimura}(1956)}]{kimura_1956}%
  \BibitemOpen
  \bibfield  {author} {\bibinfo {author} {\bibfnamefont {M.}~\bibnamefont
  {Kimura}},\ }\href@noop {} {\bibfield  {journal} {\bibinfo  {journal}
  {Biometrics}\ }\textbf {\bibinfo {volume} {12}},\ \bibinfo {pages} {57}
  (\bibinfo {year} {1956})}\BibitemShut {NoStop}%
\bibitem [{\citenamefont {Zeng}\ \emph {et~al.}(1989)\citenamefont {Zeng},
  \citenamefont {Tachida},\ and\ \citenamefont {Cockerham}}]{zeng_1989}%
  \BibitemOpen
  \bibfield  {author} {\bibinfo {author} {\bibfnamefont {Z.~B.}\ \bibnamefont
  {Zeng}}, \bibinfo {author} {\bibfnamefont {H.}~\bibnamefont {Tachida}}, \
  and\ \bibinfo {author} {\bibfnamefont {C.}~\bibnamefont {Cockerham}},\
  }\href@noop {} {\bibfield  {journal} {\bibinfo  {journal} {Genetics}\
  }\textbf {\bibinfo {volume} {11}},\ \bibinfo {pages} {977} (\bibinfo {year}
  {1989})}\BibitemShut {NoStop}%
\bibitem [{\citenamefont {Mallet}(2012)}]{mallet_2012}%
  \BibitemOpen
  \bibfield  {author} {\bibinfo {author} {\bibfnamefont {J.}~\bibnamefont
  {Mallet}},\ }\href@noop {} {\bibfield  {journal} {\bibinfo  {journal} {Evol.
  Ecol. Res.}\ }\textbf {\bibinfo {volume} {14}},\ \bibinfo {pages} {627}
  (\bibinfo {year} {2012})}\BibitemShut {NoStop}%
\bibitem [{\citenamefont {Otto}\ \emph {et~al.}(2008)\citenamefont {Otto},
  \citenamefont {Servedio},\ and\ \citenamefont {Nuismer}}]{otto_2008}%
  \BibitemOpen
  \bibfield  {author} {\bibinfo {author} {\bibfnamefont {S.~P.}\ \bibnamefont
  {Otto}}, \bibinfo {author} {\bibfnamefont {M.~R.}\ \bibnamefont {Servedio}},
  \ and\ \bibinfo {author} {\bibfnamefont {S.~L.}\ \bibnamefont {Nuismer}},\
  }\href@noop {} {\bibfield  {journal} {\bibinfo  {journal} {Genetics}\
  }\textbf {\bibinfo {volume} {179}},\ \bibinfo {pages} {2091} (\bibinfo {year}
  {2008})}\BibitemShut {NoStop}%
\bibitem [{\citenamefont {Antal}\ \emph {et~al.}(2009)\citenamefont {Antal},
  \citenamefont {Traulsen}, \citenamefont {Ohtsuki}, \citenamefont {Tarnita},\
  and\ \citenamefont {Nowak}}]{antal_2009}%
  \BibitemOpen
  \bibfield  {author} {\bibinfo {author} {\bibfnamefont {T.}~\bibnamefont
  {Antal}}, \bibinfo {author} {\bibfnamefont {A.}~\bibnamefont {Traulsen}},
  \bibinfo {author} {\bibfnamefont {H.}~\bibnamefont {Ohtsuki}}, \bibinfo
  {author} {\bibfnamefont {C.~E.}\ \bibnamefont {Tarnita}}, \ and\ \bibinfo
  {author} {\bibfnamefont {M.~A.}\ \bibnamefont {Nowak}},\ }\href@noop {}
  {\bibfield  {journal} {\bibinfo  {journal} {J. Theor. Biol.}\ }\textbf
  {\bibinfo {volume} {258}},\ \bibinfo {pages} {614} (\bibinfo {year}
  {2009})}\BibitemShut {NoStop}%
\bibitem [{\citenamefont {Hofbauer}(1981)}]{hofbauer_1981}%
  \BibitemOpen
  \bibfield  {author} {\bibinfo {author} {\bibfnamefont {J.}~\bibnamefont
  {Hofbauer}},\ }\href@noop {} {\bibfield  {journal} {\bibinfo  {journal} {J.
  Nonlinear Anal.}\ }\textbf {\bibinfo {volume} {5}},\ \bibinfo {pages} {1003}
  (\bibinfo {year} {1981})}\BibitemShut {NoStop}%
\bibitem [{\citenamefont {Bomze}(1995)}]{bomze_1995}%
  \BibitemOpen
  \bibfield  {author} {\bibinfo {author} {\bibfnamefont {I.~M.}\ \bibnamefont
  {Bomze}},\ }\href@noop {} {\bibfield  {journal} {\bibinfo  {journal}
  {Biological Cybernetics}\ }\textbf {\bibinfo {volume} {72}},\ \bibinfo
  {pages} {447} (\bibinfo {year} {1995})}\BibitemShut {NoStop}%
\bibitem [{\citenamefont {Khasminskii}\ and\ \citenamefont
  {Klebaner}(2001)}]{khasminskii_2001}%
  \BibitemOpen
  \bibfield  {author} {\bibinfo {author} {\bibfnamefont {R.~Z.}\ \bibnamefont
  {Khasminskii}}\ and\ \bibinfo {author} {\bibfnamefont {F.~C.}\ \bibnamefont
  {Klebaner}},\ }\href@noop {} {\bibfield  {journal} {\bibinfo  {journal} {Ann.
  Appl. Probab.}\ }\textbf {\bibinfo {volume} {11}},\ \bibinfo {pages} {952}
  (\bibinfo {year} {2001})}\BibitemShut {NoStop}%
\bibitem [{\citenamefont {Spagnolo}\ \emph {et~al.}(2003)\citenamefont
  {Spagnolo}, \citenamefont {Fiasconaro},\ and\ \citenamefont
  {Valenti}}]{spagnolo_2003}%
  \BibitemOpen
  \bibfield  {author} {\bibinfo {author} {\bibfnamefont {B.}~\bibnamefont
  {Spagnolo}}, \bibinfo {author} {\bibfnamefont {A.}~\bibnamefont
  {Fiasconaro}}, \ and\ \bibinfo {author} {\bibfnamefont {D.}~\bibnamefont
  {Valenti}},\ }\href@noop {} {\bibfield  {journal} {\bibinfo  {journal}
  {Fluctuation and Noise Letters}\ }\textbf {\bibinfo {volume} {3}},\ \bibinfo
  {pages} {177} (\bibinfo {year} {2003})}\BibitemShut {NoStop}%
\bibitem [{\citenamefont {Zelnik}\ \emph {et~al.}(2015)\citenamefont {Zelnik},
  \citenamefont {Solomon},\ and\ \citenamefont {Yaari}}]{zelnik_2015}%
  \BibitemOpen
  \bibfield  {author} {\bibinfo {author} {\bibfnamefont {Y.~R.}\ \bibnamefont
  {Zelnik}}, \bibinfo {author} {\bibfnamefont {S.}~\bibnamefont {Solomon}}, \
  and\ \bibinfo {author} {\bibfnamefont {G.}~\bibnamefont {Yaari}},\
  }\href@noop {} {\bibfield  {journal} {\bibinfo  {journal} {Scientific
  Reports}\ }\textbf {\bibinfo {volume} {5}},\ \bibinfo {pages} {7877}
  (\bibinfo {year} {2015})}\BibitemShut {NoStop}%
\bibitem [{\citenamefont {Noble}\ \emph {et~al.}(2011)\citenamefont {Noble},
  \citenamefont {Hastings},\ and\ \citenamefont {Fagan}}]{noble_hastings_2011}%
  \BibitemOpen
  \bibfield  {author} {\bibinfo {author} {\bibfnamefont {A.~E.}\ \bibnamefont
  {Noble}}, \bibinfo {author} {\bibfnamefont {A.}~\bibnamefont {Hastings}}, \
  and\ \bibinfo {author} {\bibfnamefont {W.~F.}\ \bibnamefont {Fagan}},\
  }\href@noop {} {\bibfield  {journal} {\bibinfo  {journal} {Phys. Rev. Lett.}\
  }\textbf {\bibinfo {volume} {107}},\ \bibinfo {pages} {228101} (\bibinfo
  {year} {2011})}\BibitemShut {NoStop}%
\bibitem [{\citenamefont {Parsons}\ \emph {et~al.}(2010)\citenamefont
  {Parsons}, \citenamefont {Quince},\ and\ \citenamefont
  {Plotkin}}]{parsons_quince_2010}%
  \BibitemOpen
  \bibfield  {author} {\bibinfo {author} {\bibfnamefont {T.~L.}\ \bibnamefont
  {Parsons}}, \bibinfo {author} {\bibfnamefont {C.}~\bibnamefont {Quince}}, \
  and\ \bibinfo {author} {\bibfnamefont {J.~B.}\ \bibnamefont {Plotkin}},\
  }\href@noop {} {\bibfield  {journal} {\bibinfo  {journal} {Genetics}\
  }\textbf {\bibinfo {volume} {185}},\ \bibinfo {pages} {1345} (\bibinfo {year}
  {2010})}\BibitemShut {NoStop}%
\bibitem [{\citenamefont {Constable}\ and\ \citenamefont
  {McKane}(2015)}]{constable_2015}%
  \BibitemOpen
  \bibfield  {author} {\bibinfo {author} {\bibfnamefont {G.~W.~A.}\
  \bibnamefont {Constable}}\ and\ \bibinfo {author} {\bibfnamefont {A.~J.}\
  \bibnamefont {McKane}},\ }\href@noop {} {\bibfield  {journal} {\bibinfo
  {journal} {Phys. Rev. Lett.}\ }\textbf {\bibinfo {volume} {114}},\ \bibinfo
  {pages} {038101} (\bibinfo {year} {2015})}\BibitemShut {NoStop}%
\bibitem [{\citenamefont {van Kampen}(2007)}]{van_Kampen_2007}%
  \BibitemOpen
  \bibfield  {author} {\bibinfo {author} {\bibfnamefont {N.~G.}\ \bibnamefont
  {van Kampen}},\ }\href@noop {} {\emph {\bibinfo {title} {Stochastic
  {P}rocesses in {P}hysics and {C}hemistry}}}\ (\bibinfo  {publisher}
  {Elsevier},\ \bibinfo {address} {Amsterdam},\ \bibinfo {year}
  {2007})\BibitemShut {NoStop}%
\bibitem [{\citenamefont {Crow}\ and\ \citenamefont
  {Kimura}(1970)}]{crow_kimura_into}%
  \BibitemOpen
  \bibfield  {author} {\bibinfo {author} {\bibfnamefont {J.~F.}\ \bibnamefont
  {Crow}}\ and\ \bibinfo {author} {\bibfnamefont {M.}~\bibnamefont {Kimura}},\
  }\href@noop {} {\emph {\bibinfo {title} {{A}n {I}ntroduction to {P}opulation
  {G}enetics {T}heory}}}\ (\bibinfo  {publisher} {The Blackburn Press},\
  \bibinfo {address} {New Jersey},\ \bibinfo {year} {1970})\BibitemShut
  {NoStop}%
\bibitem [{\citenamefont {Gardiner}(2009)}]{gardiner_2009}%
  \BibitemOpen
  \bibfield  {author} {\bibinfo {author} {\bibfnamefont {C.~W.}\ \bibnamefont
  {Gardiner}},\ }\href@noop {} {\emph {\bibinfo {title} {{H}andbook of
  {S}tochastic {M}ethods}}}\ (\bibinfo  {publisher} {Springer},\ \bibinfo
  {address} {Berlin},\ \bibinfo {year} {2009})\BibitemShut {NoStop}%
\bibitem [{\citenamefont {McKane}\ \emph {et~al.}(2014)\citenamefont {McKane},
  \citenamefont {Biancalani},\ and\ \citenamefont {Rogers}}]{mckane_BMB}%
  \BibitemOpen
  \bibfield  {author} {\bibinfo {author} {\bibfnamefont {A.~J.}\ \bibnamefont
  {McKane}}, \bibinfo {author} {\bibfnamefont {T.}~\bibnamefont {Biancalani}},
  \ and\ \bibinfo {author} {\bibfnamefont {T.}~\bibnamefont {Rogers}},\
  }\href@noop {} {\bibfield  {journal} {\bibinfo  {journal} {Bull. Math.
  Biol.}\ }\textbf {\bibinfo {volume} {76}},\ \bibinfo {pages} {895} (\bibinfo
  {year} {2014})}\BibitemShut {NoStop}%
\bibitem [{\citenamefont {Blythe}\ and\ \citenamefont
  {McKane}(2007)}]{blythe_mckane_models_2007}%
  \BibitemOpen
  \bibfield  {author} {\bibinfo {author} {\bibfnamefont {R.~A.}\ \bibnamefont
  {Blythe}}\ and\ \bibinfo {author} {\bibfnamefont {A.~J.}\ \bibnamefont
  {McKane}},\ }\href@noop {} {\bibfield  {journal} {\bibinfo  {journal} {J.
  Stat. Mech.}\ ,\ \bibinfo {pages} {P07018}} (\bibinfo {year}
  {2007})}\BibitemShut {NoStop}%
\bibitem [{\citenamefont {Kimura}(1955)}]{kimura_1955}%
  \BibitemOpen
  \bibfield  {author} {\bibinfo {author} {\bibfnamefont {M.}~\bibnamefont
  {Kimura}},\ }\href@noop {} {\bibfield  {journal} {\bibinfo  {journal}
  {Evolution}\ }\textbf {\bibinfo {volume} {9}},\ \bibinfo {pages} {419}
  (\bibinfo {year} {1955})}\BibitemShut {NoStop}%
\bibitem [{\citenamefont {Nowak}(2006)}]{nowak_2006}%
  \BibitemOpen
  \bibfield  {author} {\bibinfo {author} {\bibfnamefont {M.~A.}\ \bibnamefont
  {Nowak}},\ }\href@noop {} {\emph {\bibinfo {title} {Evolutionary Dynamics:
  exploring the equations of life}}}\ (\bibinfo  {publisher} {Harvard
  University Press},\ \bibinfo {address} {Cambridge, Massachusetts},\ \bibinfo
  {year} {2006})\BibitemShut {NoStop}%
\bibitem [{\citenamefont {Constable}\ and\ \citenamefont
  {McKane}(2014{\natexlab{a}})}]{constable_phys}%
  \BibitemOpen
  \bibfield  {author} {\bibinfo {author} {\bibfnamefont {G.~W.~A.}\
  \bibnamefont {Constable}}\ and\ \bibinfo {author} {\bibfnamefont {A.~J.}\
  \bibnamefont {McKane}},\ }\href@noop {} {\bibfield  {journal} {\bibinfo
  {journal} {Phys. Rev. E}\ }\textbf {\bibinfo {volume} {89}},\ \bibinfo
  {pages} {032141} (\bibinfo {year} {2014}{\natexlab{a}})}\BibitemShut
  {NoStop}%
\bibitem [{\citenamefont {Constable}\ and\ \citenamefont
  {McKane}(2014{\natexlab{b}})}]{constable_bio}%
  \BibitemOpen
  \bibfield  {author} {\bibinfo {author} {\bibfnamefont {G.~W.~A.}\
  \bibnamefont {Constable}}\ and\ \bibinfo {author} {\bibfnamefont {A.~J.}\
  \bibnamefont {McKane}},\ }\href@noop {} {\bibfield  {journal} {\bibinfo
  {journal} {J. Theor. Biol.}\ }\textbf {\bibinfo {volume} {358}},\ \bibinfo
  {pages} {149} (\bibinfo {year} {2014}{\natexlab{b}})}\BibitemShut {NoStop}%
\bibitem [{\citenamefont {Wiggins}(2003)}]{wiggins_2003}%
  \BibitemOpen
  \bibfield  {author} {\bibinfo {author} {\bibfnamefont {S.}~\bibnamefont
  {Wiggins}},\ }\href@noop {} {\emph {\bibinfo {title} {Introduction to
  {A}pplied {N}onlinear {D}ynamical {S}ystems and {C}haos}}}\ (\bibinfo
  {publisher} {Springer},\ \bibinfo {address} {New York},\ \bibinfo {year}
  {2003})\BibitemShut {NoStop}%
\bibitem [{\citenamefont {Gokhale}\ and\ \citenamefont
  {Traulsen}(2010)}]{gokhale_2010}%
  \BibitemOpen
  \bibfield  {author} {\bibinfo {author} {\bibfnamefont {C.~S.}\ \bibnamefont
  {Gokhale}}\ and\ \bibinfo {author} {\bibfnamefont {A.}~\bibnamefont
  {Traulsen}},\ }\href@noop {} {\bibfield  {journal} {\bibinfo  {journal}
  {Proc. Natl. Acad. Sci}\ }\textbf {\bibinfo {volume} {107}},\ \bibinfo
  {pages} {5500} (\bibinfo {year} {2010})}\BibitemShut {NoStop}%
\bibitem [{\citenamefont {Baxter}\ \emph {et~al.}(2007)\citenamefont {Baxter},
  \citenamefont {Blythe},\ and\ \citenamefont {McKane}}]{baxter_2007}%
  \BibitemOpen
  \bibfield  {author} {\bibinfo {author} {\bibfnamefont {G.~J.}\ \bibnamefont
  {Baxter}}, \bibinfo {author} {\bibfnamefont {R.~A.}\ \bibnamefont {Blythe}},
  \ and\ \bibinfo {author} {\bibfnamefont {A.~J.}\ \bibnamefont {McKane}},\
  }\href@noop {} {\bibfield  {journal} {\bibinfo  {journal} {Math. Biosci.}\
  }\textbf {\bibinfo {volume} {209}},\ \bibinfo {pages} {124} (\bibinfo {year}
  {2007})}\BibitemShut {NoStop}%
\bibitem [{\citenamefont {Abramowitz}\ and\ \citenamefont
  {Stegun}(1965)}]{handbook_1965}%
  \BibitemOpen
  \bibinfo {editor} {\bibfnamefont {M.}~\bibnamefont {Abramowitz}}\ and\
  \bibinfo {editor} {\bibfnamefont {I.~A.}\ \bibnamefont {Stegun}},\ eds.,\
  \href@noop {} {\emph {\bibinfo {title} {{H}andbook of {M}athematical
  {F}unctions}}}\ (\bibinfo  {publisher} {Dover Publications},\ \bibinfo
  {address} {New York},\ \bibinfo {year} {1965})\BibitemShut {NoStop}%
\bibitem [{\citenamefont {Erd\'{e}lyi}(1953)}]{HTFS_1953}%
  \BibitemOpen
  \bibinfo {editor} {\bibfnamefont {A.}~\bibnamefont {Erd\'{e}lyi}},\ ed.,\
  \href@noop {} {\emph {\bibinfo {title} {{H}igher {T}ranscendental
  {F}unctions: Vol II}}}\ (\bibinfo  {publisher} {McGraw-Hill},\ \bibinfo
  {address} {New York},\ \bibinfo {year} {1953})\BibitemShut {NoStop}%
\bibitem [{\citenamefont {Doering}\ \emph {et~al.}(2006)\citenamefont
  {Doering}, \citenamefont {Sargsyan},\ and\ \citenamefont
  {Sander}}]{doering_2006}%
  \BibitemOpen
  \bibfield  {author} {\bibinfo {author} {\bibfnamefont {C.~R.}\ \bibnamefont
  {Doering}}, \bibinfo {author} {\bibfnamefont {K.~V.}\ \bibnamefont
  {Sargsyan}}, \ and\ \bibinfo {author} {\bibfnamefont {L.~M.}\ \bibnamefont
  {Sander}},\ }\href@noop {} {\bibfield  {journal} {\bibinfo  {journal}
  {Multiscale Model. Simul}\ }\textbf {\bibinfo {volume} {3}},\ \bibinfo
  {pages} {283} (\bibinfo {year} {2006})}\BibitemShut {NoStop}%
\bibitem [{\citenamefont {Constable}\ \emph {et~al.}(2016)\citenamefont
  {Constable}, \citenamefont {Rogers}, \citenamefont {McKane},\ and\
  \citenamefont {Tarnita}}]{constable_2016}%
  \BibitemOpen
  \bibfield  {author} {\bibinfo {author} {\bibfnamefont {G.~W.~A.}\
  \bibnamefont {Constable}}, \bibinfo {author} {\bibfnamefont {T.}~\bibnamefont
  {Rogers}}, \bibinfo {author} {\bibfnamefont {A.~J.}\ \bibnamefont {McKane}},
  \ and\ \bibinfo {author} {\bibfnamefont {C.~E.}\ \bibnamefont {Tarnita}},\
  }\href@noop {} {\bibfield  {journal} {\bibinfo  {journal} {Proc. Natl. Acad.
  Sci.}\ }\textbf {\bibinfo {volume} {113}},\ \bibinfo {pages} {E4745}
  (\bibinfo {year} {2016})}\BibitemShut {NoStop}%
\bibitem [{\citenamefont {Lin}\ \emph {et~al.}(2015)\citenamefont {Lin},
  \citenamefont {Kim},\ and\ \citenamefont {Doering}}]{lin_mig_1}%
  \BibitemOpen
  \bibfield  {author} {\bibinfo {author} {\bibfnamefont {Y.~T.}\ \bibnamefont
  {Lin}}, \bibinfo {author} {\bibfnamefont {H.}~\bibnamefont {Kim}}, \ and\
  \bibinfo {author} {\bibfnamefont {C.~R.}\ \bibnamefont {Doering}},\
  }\href@noop {} {\bibfield  {journal} {\bibinfo  {journal} {J. Math. Biol.}\
  }\textbf {\bibinfo {volume} {70}},\ \bibinfo {pages} {647} (\bibinfo {year}
  {2015})}\BibitemShut {NoStop}%
\bibitem [{\citenamefont {Assaf}\ and\ \citenamefont
  {Meerson}(2010)}]{assaf_2010}%
  \BibitemOpen
  \bibfield  {author} {\bibinfo {author} {\bibfnamefont {M.}~\bibnamefont
  {Assaf}}\ and\ \bibinfo {author} {\bibfnamefont {B.}~\bibnamefont
  {Meerson}},\ }\href@noop {} {\bibfield  {journal} {\bibinfo  {journal} {Phys.
  Rev. E}\ }\textbf {\bibinfo {volume} {81}},\ \bibinfo {pages} {021116}
  (\bibinfo {year} {2010})}\BibitemShut {NoStop}%
\bibitem [{\citenamefont {Vogl}\ and\ \citenamefont
  {Clemente}(2012)}]{vogl_2012}%
  \BibitemOpen
  \bibfield  {author} {\bibinfo {author} {\bibfnamefont {C.}~\bibnamefont
  {Vogl}}\ and\ \bibinfo {author} {\bibfnamefont {F.}~\bibnamefont
  {Clemente}},\ }\href {\doibase http://dx.doi.org/10.1016/j.tpb.2012.01.001}
  {\bibfield  {journal} {\bibinfo  {journal} {Theor. Popul. Biol.}\ }\textbf
  {\bibinfo {volume} {81}},\ \bibinfo {pages} {197 } (\bibinfo {year}
  {2012})}\BibitemShut {NoStop}%
\bibitem [{\citenamefont {Maddamsetti}\ \emph {et~al.}(2015)\citenamefont
  {Maddamsetti}, \citenamefont {Lenski},\ and\ \citenamefont
  {Barrick}}]{maddamsetti_2015}%
  \BibitemOpen
  \bibfield  {author} {\bibinfo {author} {\bibfnamefont {R.}~\bibnamefont
  {Maddamsetti}}, \bibinfo {author} {\bibfnamefont {R.~E.}\ \bibnamefont
  {Lenski}}, \ and\ \bibinfo {author} {\bibfnamefont {J.~E.}\ \bibnamefont
  {Barrick}},\ }\href@noop {} {\bibfield  {journal} {\bibinfo  {journal}
  {Genetics}\ }\textbf {\bibinfo {volume} {200}},\ \bibinfo {pages} {619}
  (\bibinfo {year} {2015})}\BibitemShut {NoStop}%
\bibitem [{\citenamefont {Huang}\ \emph {et~al.}(2012)\citenamefont {Huang},
  \citenamefont {Haubold}, \citenamefont {Hauert},\ and\ \citenamefont
  {Traulsen}}]{antal_2012}%
  \BibitemOpen
  \bibfield  {author} {\bibinfo {author} {\bibfnamefont {W.}~\bibnamefont
  {Huang}}, \bibinfo {author} {\bibfnamefont {B.}~\bibnamefont {Haubold}},
  \bibinfo {author} {\bibfnamefont {C.}~\bibnamefont {Hauert}}, \ and\ \bibinfo
  {author} {\bibfnamefont {A.}~\bibnamefont {Traulsen}},\ }\href@noop {}
  {\bibfield  {journal} {\bibinfo  {journal} {Nature Comms.}\ }\textbf
  {\bibinfo {volume} {3}},\ \bibinfo {pages} {919} (\bibinfo {year}
  {2012})}\BibitemShut {NoStop}%
\bibitem [{\citenamefont {Gokhale}\ \emph {et~al.}(2013)\citenamefont
  {Gokhale}, \citenamefont {Papkou}, \citenamefont {Traulsen},\ and\
  \citenamefont {Schulenburg}}]{gokhale_2013}%
  \BibitemOpen
  \bibfield  {author} {\bibinfo {author} {\bibfnamefont {C.~S.}\ \bibnamefont
  {Gokhale}}, \bibinfo {author} {\bibfnamefont {A.}~\bibnamefont {Papkou}},
  \bibinfo {author} {\bibfnamefont {A.}~\bibnamefont {Traulsen}}, \ and\
  \bibinfo {author} {\bibfnamefont {H.}~\bibnamefont {Schulenburg}},\
  }\href@noop {} {\bibfield  {journal} {\bibinfo  {journal} {BMC Evolutionary
  Biology}\ }\textbf {\bibinfo {volume} {13}},\ \bibinfo {pages} {254}
  (\bibinfo {year} {2013})}\BibitemShut {NoStop}%
\bibitem [{\citenamefont {Constable}(2015)}]{constable_thesis}%
  \BibitemOpen
  \bibfield  {author} {\bibinfo {author} {\bibfnamefont {G.~W.~A.}\
  \bibnamefont {Constable}},\ }\href@noop {} {\emph {\bibinfo {title} {{F}ast
  {V}ariables in {S}tochastic {P}opulation {D}ynamics}}},\ Springer Theses
  Series\ (\bibinfo  {publisher} {Springer},\ \bibinfo {address} {Heidelberg},\
  \bibinfo {year} {2015})\BibitemShut {NoStop}%
\end{thebibliography}

%

\end{document}